\documentclass[preprint, 3p]{elsarticle}
\journal{NLM}
\usepackage{hyperref}
\hypersetup{
 pdftitle    = {pdftitle},
 pdfsubject  = {pdfsubject},
 pdfauthor   = {pdfauthor},
 pdfkeywords = {pdfkexwords},
 pdfborder   = 0 0 0,
 plainpages  = false,
 bookmarksnumbered = true
}
\usepackage[usenames,dvipsnames]{xcolor}
\usepackage{graphicx}           
\usepackage{amsmath}
\usepackage{amsfonts}
\usepackage{amssymb}
\usepackage{xspace}
\usepackage{color}
\usepackage{relsize}
\usepackage{float}
\graphicspath{}
\newcommand{\nc}{\newcommand}
\nc{\rnc}{\renewcommand}
\nc{\bs}{\boldsymbol}
\rnc{\matrix}[2]{\left[\!\!\begin{array}{#1}
	#2\end{array}\!\!\right]}
\rnc{\vector}[1]{\matrix{c}{#1}}
\nc{\mm}[1]{\boldsymbol{#1}}
\nc{\mms}[1]{\boldsymbol{#1}}
\nc{\mrm}[1]{\mathrm{#1}} 
\nc{\real}[1]{\Re\lbrace #1 \rbrace}
\nc{\imag}[1]{\Im\lbrace #1 \rbrace}
\nc{\realnumber}{\mathbb{R}} 
\nc{\complexnumber}{\mathbb{C}} 
\nc{\dd}{\mathrm{d}}
\nc{\ii}{\mathrm{i}}
\nc{\ee}{\mathrm{e}}
\nc{\inv}{^{-1}} 
\nc{\herm}{^{\mathrm H}}
\nc{\tra}{^{\mathrm T}}
\nc{\conj}[1]{ \overline{#1} }
\nc{\normal}{\mathrm n}
\nc{\tangential}{\mathrm t}
\nc{\kn}{{k_{\normal}}}
\nc{\kt}{{k_{\tangential}}}
\nc{\exa}{\hat{\ddot w}_0} 
\nc{\exaval}[1]{\hat{\ddot w}_0=#1 \mrm{m/s}^2} 
\nc{\ndof}{N_{\mathrm{dof}}} 
\nc{\nsens}{N_{\mathrm{sens}}} 
\nc{\nmod}{N_{\mathrm{mode}}} 
\nc{\bphi}{\bs\phi} 
\nc{\wrt}{with respect to\xspace} 
\definecolor{myblack}{rgb}{0.2,0.2,0.2}
\definecolor{mygreen}{rgb}{0,0.8110,0.3149}
\definecolor{myred}{rgb}{0.9922,0.2275,0.3961}
\definecolor{myblue}{rgb}{0,0.4,0.8}
\nc{\Pone}{P_1 \textcolor{myblue}{\pmb \times}}
\nc{\Ptwo}{P_2 \textcolor{myblack}{\pmb \times}}
\nc{\Pthree}{P_3 \textcolor{mygreen}{\pmb \square}}
\nc{\Pfour}{P_4 \textcolor{myred}{\pmb \square}}
\nc{\sSIM}{sSIM\xspace}
\nc{\COMMENT}[1]{#1}
\nc{\MOD}[1]{#1}
\nc{\ie}{i.\,e.\xspace}
\nc{\eg}{e.\,g.\xspace}
\nc{\cf}{cf.\xspace}
\nc{\myquote}[1]{`#1'}
\nc{\etal}{et al.\xspace}
\nc{\fabstand}{\,}
\nc{\fp}{\fabstand.}
\nc{\fk}{\fabstand,}
\rnc{\phi}{\varphi}
\nc{\slab}[1]{\label{sec:#1}}
\nc{\flab}[1]{\label{fig:#1}}
\nc{\tlab}[1]{\label{tab:#1}}
\nc{\elab}[1]{\label{eq:#1}}
\nc{\sect}[2]{\section{#1 \slab{#2}}}
\nc{\ssect}[2]{\subsection{#1 \slab{#2}}}
\nc{\sssect}[2]{\subsubsection{#1 \slab{#2}}}
\nc{\tab}[5][tbh]{\begin{table}[#1]\centering\caption{#4\label{tab:#5}}\begin{tabular}{#2}\hline #3 \\ \hline\end{tabular}\end{table}}
\newcommand{\onefig}[4]{%
\begin{figure}[h!]
	\begin{center}
		\includegraphics[width=#1\textwidth]{#2}		
		\caption{#3}
		\flab{#4}
	\end{center}
\end{figure}
}
\newcommand{\onelargefig}[4]{%
\begin{figure}[!]
	\begin{center}
		\includegraphics[width=#1\textwidth]{#2}		
		\caption{#3}
		\flab{#4}
	\end{center}
\end{figure}
}
\nc{\e}[2]{\begin{equation} #1 \label {eq:#2} \end{equation}}
\nc{\est}[1]{\begin{equation*} #1 \end{equation*}}
\nc{\ea}[1]{
\begin{eqnarray}
#1\end{eqnarray}}
\nc{\east}[1]{
\begin{eqnarray*}
#1 \end{eqnarray*}}
\nc{\fref}[1]{{Fig.~\ref{fig:#1}}}
\nc{\frefo}[1]{{\ref{fig:#1}}}
\nc{\frefs}[1]{{Figs.~\ref{fig:#1}}}
\nc{\frefsf}[2]{{Fig.~\ref{fig:#1}{#2}}} 
\nc{\tref}[1]{{Tab.~\ref{tab:#1}}}
\nc{\trefo}[1]{{\ref{tab:#1}}}
\nc{\trefs}[1]{{Tab.~\ref{tab:#1}}}
\nc{\eref}[1]{{Eq.~(\ref{eq:#1})}}
\nc{\erefo}[1]{(\ref{eq:#1})}
\nc{\erefs}[1]{{Eqs.~(\ref{eq:#1})}}
\nc{\sref}[1]{{Section~\ref{sec:#1}}}
\nc{\srefo}[1]{\ref{sec:#1}}
\nc{\srefs}[1]{{Sections~\ref{sec:#1}}}
\nc{\ssref}[1]{{Section~\ref{sec:#1}}} 
\nc{\ssrefo}[1]{\ref{sec:#1}} 
\nc{\ssrefs}[1]{{Sections~\ref{sec:#1}}} 
\nc{\aref}[1]{{{\ref{asec:#1}}}}
\nc{\arefo}[1]{{\ref{asec:#1}}}
\nc{\arefs}[1]{{{Appendices~\ref{asec:#1}}}}

\makeindex             

\begin{document}

\begin{frontmatter}
\title{On the locomotion of the slider within a self-adaptive beam-slider system}
\author{
Florian Müller$^1$,
Malte Krack$^2$
}
\address{$^1$ Carl Zeiss SMT GmbH, GERMANY}
\address{$^2$ University of Stuttgart, GERMANY}

\begin{abstract}
A beam-slider system is considered whose passive self-adaption relies on an intricate locomotion process involving both frictional and unilateral contact.
The system also exploits geometric nonlinearity to achieve broadband efficacy.
The dynamics of the system take place on three distinct time scales:
On the fast time scale of the harmonic base excitation are the vibrations and the locomotion cycle.
On the slow time scale, the slider changes its position along the beam, and the overall vibration level varies.
Finally, on an intermediate time scale, strong modulations of the vibration amplitude may take place.
In the present work, first, an analytical approximation of the beam's response on the slow time scale is derived as function of the slider position, which is a crucial prerequisite for identifying the main drivers of the slider's locomotion.
Then, the most important forms of locomotion are described and approximations of their individual contribution to the overall slider transport are estimated.
Finally, the theoretical results are compared against numerical results obtained from an experimentally validated model.
\end{abstract}

\begin{keyword}
vibration-induced sliding; locomotion; nonlinear dynamics; multiple scales; self-tuning 
\end{keyword}

\end{frontmatter}

\section{Introduction}
The beam-slider system considered in the present work is an example of an intelligent use of strong nonlinearity.
The nonlinear bending-stretching coupling of the clamped-clamped beam permits the system to have broadband efficacy, which is favorable for the use in vibration mitigation or vibration energy harvesting applications.
As will be shown in this work, dry friction and free play nonlinearity are essential for the fully passive rearrangement of the slider along the beam, enabling to switch from low to high vibrations.
The following paragraph describes the setup and typical behavior of the considered system, and is essentially adopted from \cite{Muller.2023}.
\onefig{1}{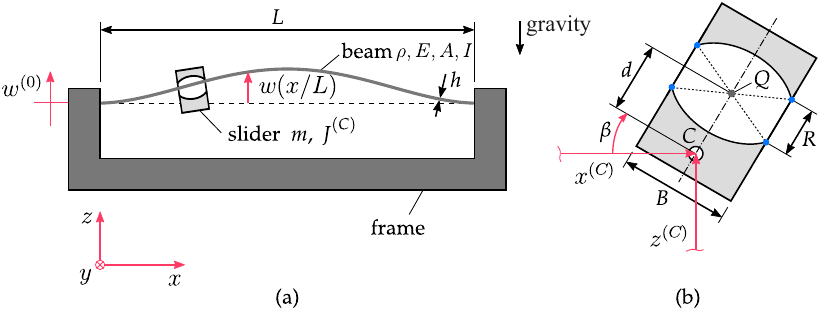}{Schematic of self-adaptive system: (a) two-dimensional model of clamped–clamped beam with attached slider, (b) slider detail.
}{SAS}
\onefig{1}{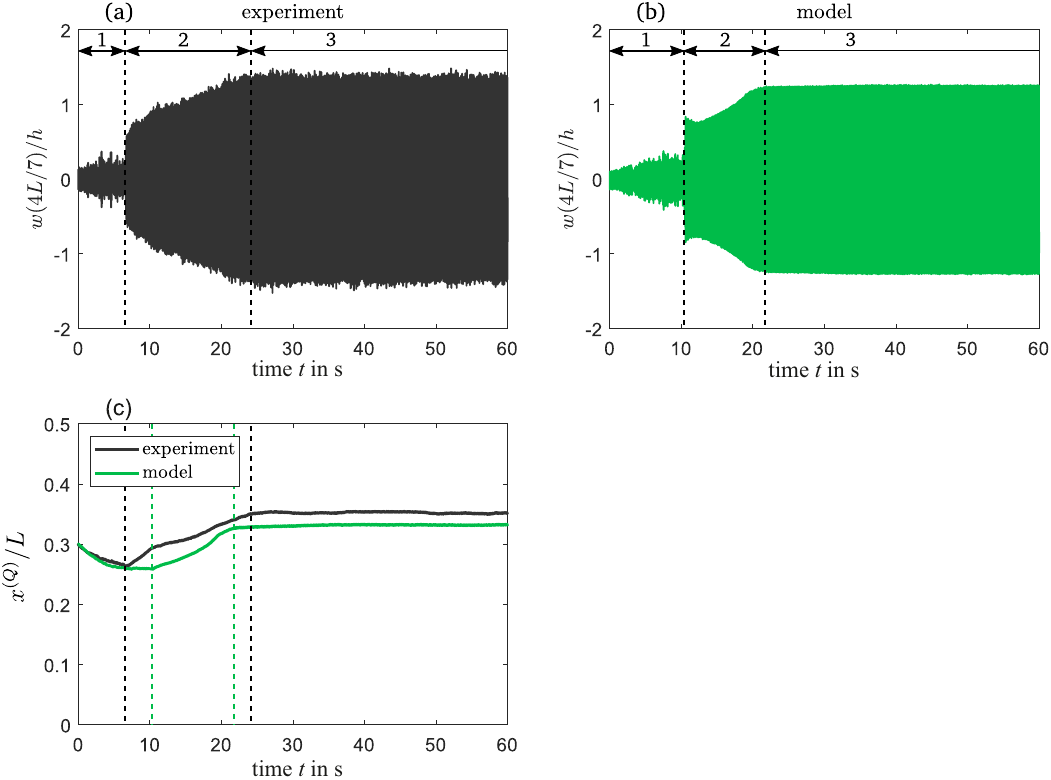}{Illustration of the signature move and confrontation of model and experiment: (a) and (b) beam vibration level vs. time; (c) slider location vs. time.}{signatureMoveintro}
\\
Consider the beam-slider system depicted in \fref{SAS}.
The system consists of a slender beam, clamped at both ends to a stiff frame, and an essentially rigid mass (slider) that is attached to the beam.
The slider's gap size $R$ is slightly larger than the beam's thickness $h$, so that there is a clearance between beam and slider.
Under harmonic base excitation in the range around the lowest-frequency bending mode, this system shows self-adaptive behavior in experiments carried out independently by different research groups \cite{Miller.2013,Yu.2020,Shin.2020,Aboulfotoh.2017,Muller.2023}.
The system was observed to adapt itself in such a way that it achieves and maintains high vibrations in a wide band of excitation frequencies.
After initially small vibrations, the slider moved to a certain position and the vibration level increased substantially.
Hereby a \emph{signature move} was observed, which is shown in~\fref{signatureMoveintro}, and is described in the following.
Also, a video of the signature move is available here: \url{https://www.youtube.com/watch?v=qSy8ccbOgn8}.
While the system vibrates initially at small level, the slider moves slowly towards the clamping.
At a certain point, the vibrations jump to a higher level and the slider turns back towards the beam's center.
This movement goes along with a further increase of the vibration level.
Before reaching the beam's center, the slider stops and large vibrations are maintained at steady state.
As this behavior occurs over a broad band of excitation frequencies, the described beam-slider system received considerable attention in the vibration energy harvesting community \cite{Aboulfotoh.2017,Gregg.2014,Shin.2020}.
Variants involving a cantilever beam/plate instead of a doubly-clamped one \cite{Yu.2020,Bukhari.2020,Lihua.2020} 
or arrangements involving multiple cantilevers \cite{Staaf.2018} 
have also been proposed for broadband energy harvesting.
\\
A relatively small amount of research effort addresses the theoretical understanding of the system, compared to studies exploring the behavior experimentally or using the system for energy harvesting.
Yu \etal \cite{Yu.2019} were the first to explain the co-existing vibration states, the amplitude jumps and the broadband efficacy by the nonlinear bending-stretching coupling of the clamped-clamped beam.
Thomsen \cite{Thomsen.1996,Miranda.1998} was one of the first to explain the movement of a sliding mass along a vibrating string/beam by what we will refer to as \emph{slope-induced sliding}.
Here, the host structure's transverse acceleration, projected onto the rotated beam axis, is postulated as main driver for the movement of the attached mass, $\partial^2 x^{(Q)}/\partial t^2 \approx - (\partial^2(w+w^{(0)})/\partial t^2~\partial w/\partial x)|_{x^{(Q)}/L}$.
This expression will be derived later, and it will be discussed that this can only explain the movement of the slider towards anti-nodes of the transverse displacement (where the rotation is zero); \ie, the last part of the movement during the signature move.
Pillatsch \etal \cite{Pillatsch.2013} observed in their experiment that the movement of the slider sensitively depends on the clearance between slider and beam.
Krack \etal \cite{Krack.2017b} were the first include the clearance in their model and demonstrated numerically that both unilateral contact and dry friction are crucial to reproduce the locomotion of the slider away from the beam's center.
That model was later extended to account for the beam's bending-stretching coupling by Müller and Krack \cite{Muller.2020}, and the numerical simulations were validated experimentally \cite{Muller.2023}.
Very good qualitative and quantitative agreement was achieved; representative results are shown in \fref{signatureMoveintro}.
\MOD{The simulation code, which reproduces experimental results from \cite{Muller.2023}, which are also publicly accessible, has been made available online at \url{https://github.com/maltekrack/simSAS}.}
In addition to the signature move, other types of behavior are also possible: trivial adaption (monotonous movement towards the beam's center), a relaxation oscillation (recurrent back-and-forth movement with intermediate amplitude jumps), rearrangement leading to low vibrations, and inactivity (no slider movement), for appropriate excitation conditions and initial slider position \cite{Gregg.2014}.
The numerical simulations showed very good agreement for all these types of behavior \cite{Muller.2023}.
However, the physical mechanisms that actually cause the locomotion are far from being understood.
\\
The purpose of the present work is to gain a deep theoretical understanding of the slider's locomotion along the vibrating beam.
First, the system dynamics on the slow time scale is analytically investigated in \sref{sSIM}, which will turn out to yield vital information for understanding the locomotion.
Then, the relevant forms of locomotion are described and their relative importance in the different phases of the signature move are estimated.
Finally, that theory is validated using numerical simulations obtained using the experimentally validated model from \cite{Muller.2023}.
Parts of the present work are content of a doctoral thesis \cite{Muller.thesis}.

\section{Super-Slow Invariant Manifold\label{sec:sSIM}}
In this section, the focus is placed on the system dynamics on the (super-)slow time scale.
Of particular interest is the relation between the beam's vibration level and the slider location, for given excitation frequency and level.
This relation is referred to as \emph{Super-Slow Invariant Manifold (\sSIM)}.
First, a reduced-order model of the beam with attached mass and inertia effects of the slider is derived.
By truncating to a single (spatial) mode and a single (temporal) harmonic, an analytical solution for the \sSIM is obtained.
The knowledge of the modal deflection shape, the phase lag between the beam's elastic and the imposed base displacement, and the typical orders of magnitude of the vibration properties is a fundamental prerequisite for understanding the locomotion.
Further, the question is addressed whether the self-adaptive system typically operates under or reaches the condition of resonance.
Finally, the error induced by the simplifications (ideal boundary conditions, rigidly attached slider) made to derive the analytical solution is assessed.

\subsection{Reduced-order model of the beam with rigidly attached slider}
The most recent prototype of the self-adaptive system has a slider that is about three times heavier than the beam \cite{Muller.2023}.
This will be seen to considerably affect the fundamental natural frequency, but only moderately affect the modal deflection shape.
Also, the finite rotational clamping stiffness of the beam was found to cause an about $5~\%$ lower fundamental natural frequency (compared to the case with perfect clamping).
The idea of the following analytical investigation is to focus on the most important effects only.
Therefore, the beam model is truncated to the fundamental mode, where the modal deflection shape $\phi(\xi)$ with $\xi=x/L$ is simply obtained under ideal (clamped-clamped) boundary conditions and without the effects of the attached slider.
Further, Euler-Bernoulli theory is used, applied to a straight and non-preloaded beam with a uniform mass density $\rho$, Young's modulus $E$ and rectangular cross section (area $A$, area moment of inertia w.r.t. bending $I$).
The corresponding mode shape is
%
\ea{
\phi(\xi) = \frac{\sin\lambda + \sinh\lambda}{\cos\lambda-\cosh\lambda} \left(~\sin\left(\lambda\xi\right)~-~\sinh\left(\lambda\xi\right)~\right) + \cos\left(\lambda\xi\right)-\cosh\left(\lambda\xi\right)\fp \label{eq:modeshape}
}
Herein, $\lambda$ is the smallest root of the frequency equation $\cos\lambda\cosh\lambda-1=0$, $\lambda\approx 4.730$.
The associated modal angular frequency is $\omega = \sqrt{\frac{EI}{\rho AL^4}}\lambda^2$.
As modal coordinate, the transverse displacement at the beam's center is used, normalized by its length, $q= w(1/2)/L$.
The time dependence is not explicitly noted here an in the following for readability.
Thus, one can express the normalized transverse displacement as continuous function of $x$ as $w(x/L)/L\approx \phi(x/L)/\phi(1/2)q$.
Modal truncation can be interpreted as requiring that the residual, obtained by substituting this ansatz into the governing partial differential equation, is orthogonal to the base function $\phi$ (Galerkin condition).
The actual partial differential equation, derived and presented in \cite{Muller.2020,Muller.2023}, is not repeated here for brevity.
The resulting ordinary differential equation is
\ea{
\left(1+\mu\right)q_{\tau\tau} + 2D q_\tau + q + \kappa q^3 &=& -\gamma \frac{w_{\tau\tau}^{(0)}}{L} \fk \label{eq:beamROM} \\
\mu(s) &=& \frac{m}{\rho AL}\phi^2(s) + \frac{J^{(C)}+md^2}{\rho AL^4}\phi_{\xi}^2\left(s\right)\fk \label{eq:mu} \\
\kappa &=& \frac{k_{\mathrm{ax}}}{\frac{EI}{L^3}\lambda^4}\frac12 \left(\frac{\int\limits_0^1\phi_\xi^2\left(\xi\right)\dd\xi}{\phi(\frac12)}\right)^2 \fk \label{eq:kappa} \\
\gamma(s) &=& \phi\left(\frac12\right)\left(~\int\limits_0^1 \phi\left(\xi\right)\dd\xi + \frac{m}{\rho AL}\phi\left(s\right)~\right) \fk \label{eq:gamma} \\
\frac{1}{k_{\mathrm{ax}}} &=& \frac{1}{\frac{EA}{L}} + \frac{1}{k_{\mathrm t}} \fp \label{eq:kax}
}
${\square}_{\tau}$ and $\square_{\xi}$ denote partial derivatives, $\square_\tau=\partial\square/\partial\tau$,
$\square_{\xi}=\partial\square/\partial \xi$, where
$\tau$ is the normalized time $\tau = \omega t$.
Further, $m$ is the slider's mass, $J^{(C)}$ is the slider's rotational moment of inertia with respect to the center of mass $C$,
and $s=x^{(Q)}/L$ is the normalized horizontal position of slider's geometric center $Q$ (\cf~\fref{SAS}b).
With the distance $d$ from $Q$ to $C$, one has $J^{(Q)}=J^{(C)}+md^2$.
The modal damping ratio is denoted as $D$, the beam's stretching stiffness is $EA/L$, and the axial stiffness of the clamping is $k_{\mathrm t}$.
\eref{mu} and \eref{gamma} express that the slider's inertia properties contribute to both the effective modal mass, and the excitation by the imposed acceleration $w_{\tau\tau}^{(0)}$.
Bending-stretching coupling leads to a cubic spring term $\kappa q^3$ in \eref{beamROM}.
The dimensionless coefficient $\kappa$ scales with the ratio between the axial stiffness $k_{\mathrm{ax}}$ and the modal bending stiffness $EI/L^3\lambda^4$.
Herein, $k_{\mathrm{ax}}$ results from the serial arrangement of the the beam's static stretching stiffness $EA/L$ and the horizontal clamping stiffness $k_{\mrm t}$ (\eref{kax}).

\subsection{Approximation of the \sSIM via single-term Harmonic Balance}
Harmonic base motion $w^{(0)} = \hat w^{(0)}\cos\left(\Omega t\right)$ with amplitude $\hat w^{(0)}\in \mathbb R_+$ and angular frequency $\Omega$ is considered.
Near primary resonance, it is justified to truncate the response to a single harmonic, $q\approx \hat q\cos\left(\Omega t + \theta\right)$.
Herein, $\hat q\in\mathbb R_+$ is the amplitude and $\theta\in [-\pi,\pi]$ is the phase lag.
Substituting this ansatz into \eref{beamROM}, and requiring that the (time-domain) residual is orthogonal to the fundamental Fourier base functions yields the complex single-term Harmonic Balance equation:
\ea{
\left(-\left(1+\mu\right)\left(\frac{\Omega}{\omega}\right)^2 + 2D\ii\frac{\Omega}{\omega} + 1 + \frac{3}{4}\kappa \hat q^2\right)\hat q\ee^{\ii\theta} &=& \gamma \left(\frac{\Omega}{\omega}\right)^2\frac{\hat w^{(0)}}{L}\fk \label{eq:SHB} \\
\frac{\tilde\omega}{\omega}(s,\hat q) &=& \sqrt{\frac{1+\frac34\kappa\hat q^2}{1+\mu(s)}} \fp \label{eq:SHBomtil}
}
Herein, $\ii=\sqrt{-1}$ is the imaginary unit.
To evaluate the time derivatives, it was used that $\Omega t = \Omega/\omega \tau$.
The effective natural frequency $\tilde\omega$ depends on the vibration amplitude $\hat q$ and the slider position $s$ (\eref{SHBomtil}).
As shown in \cite{Muller.2023}, both parameters have crucial effects.
More specifically, attaching the slider at the beam's center, reduces the modal frequency by $64~\%$ for the parameters listed in \tref{params}.
For a vibration amplitude of $\hat q=h/L$ ($\hat q=2h/L$), the modal frequency increases by $12~\%$ ($42~\%$) compared to the linear case. 
\tab[htb]{llll}{
Quantity & Symbol & Value & Unit \\
\hline
free length (beam) & $L$ & 140 & $\mrm{mm}$ \\
thickness (beam) & $h$ & 1 & $\mrm{mm}$ \\
density (beam) & $\rho$ & $7683$ & $\mrm{kg}/\mrm{m}^3$ \\
mass (beam, free length) & $\rho AL$ & $15.1$ & $\mrm g$ \\
Young's modulus (beam) & $E$ & $21.0$ & $\mrm{GPa}$ \\
mass (slider) & $m$ & $46.2$ & $\mrm{g}$ \\
rotary inertia w.r.t. center of mass (slider) & $J^{(C)}$ & $3.15$ & $\mrm{kg}\mrm{mm}^2$ \\
distance between contacts (on same side, slider) & $B$ & $10$ & $\mrm{mm}$ \\
distance geometric center to center of mass (slider) & $d$ & $4.1$ & $\mrm{mm}$ \\
relative gap size & $R/h$ & $1.05$ & $-$ \\
axial clamping stiffness & $k_{\mrm t}$ & $1.93\cdot10^7$ & $\mrm{N}/\mrm{m}$ \\
fundamental modal frequency (beam, no slider) & $\omega/(2\pi)$ & $260$ & $\mrm{Hz}$ \\
fundamental modal damping ratio (beam, no slider) & $D$ & $0.1$ & $\%$ \\
friction coefficient (beam-slider) & $-$ & $0.2$ & $-$
}{Parameters of the self-adaptive system in \cite{Muller.2023}. The properties $A$ and $I$ of the rectangular cross section can be uniquely determined by the listed parameters. The reported $\omega$ contains the effect of the finite rotational clamping stiffness, as identified in \cite{Muller.2023}.}{params}
%
%
\onefig{1}{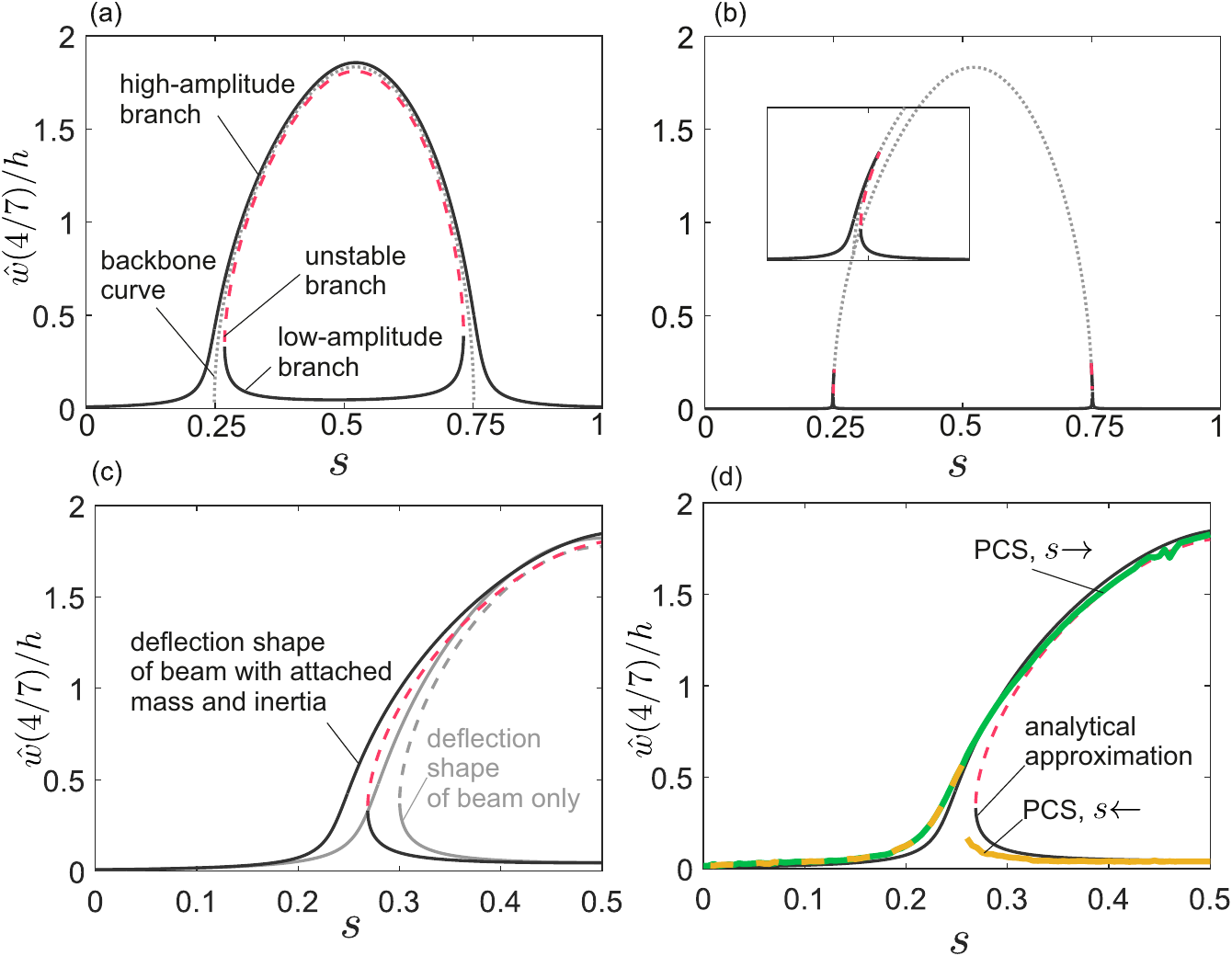}{Super-Slow Invariant Manifold (\sSIM): (a) overview of analytical approximation for parameters where self-adaptive behavior is expected; (b) the case of much lower excitation level; (c) approximation obtained with modal deflection shape of beam only vs. of beam with attached slider; (d) effect of dynamic contact interactions with slider. PCS: Pseudo-Constrained Slider.
The analytical approximation in (a), (b) and (d) is obtained with the deflection shape of the beam with attached slider.
}{sSIM}
\onefig{.5}{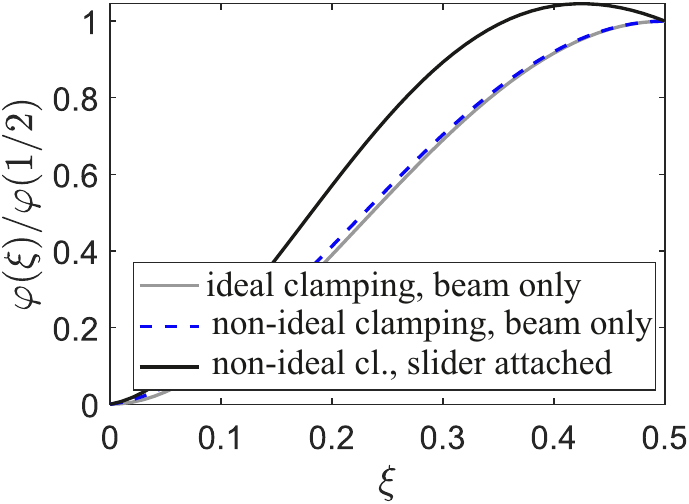}{Effect of non-ideal clamping and slider attached at $s=0.27$ on modal deflection shape.}{modalDeflectionShape}
\\
By taking the magnitude squared on both sides of \eref{SHB}, one can eliminate the unknown $\theta$, and one obtains a cubic polynomial in $\hat q^2$.
Depending on $s$, there are one, two or three real roots for $\hat q^2$.
By analyzing how the dissipated and the supplied energy per period change with the amplitude $\hat q$, one can reason the stability of the corresponding periodic limit states. 
It turns out that the limit state is stable if only one root exists \cite{Muller.thesis}.
If three roots exist, the one with intermediate amplitude is unstable, while the upper and the lower are stable.
This explains the form of the typical \sSIM depicted in \fref{sSIM}a, and the stability of its branches.
Here, the frequency and the level of the excitation are kept fixed, while the dependence of the solution $\hat q$ on $s$ is depicted.
Instead of $\hat q$, the variable $\hat w(4/7)/h$ is shown to be consistent with \fref{signatureMoveintro}; a value of $1$ means that the displacement at $x=4/7L$ is equal to its thickness.
\MOD{
The sub-figures \fref{sSIM}b-d are explained later in \ssref{sSIMvalidation}.
}
\\
\MOD{
Having established the \sSIM as depicted in \fref{sSIM}, permits to gain deeper insight into the super-slow dynamics of the signature move.
In the example given in \fref{signatureMoveintro}, one} departs near $s=0.3$ on the low-amplitude branch, where the slider moves away from the beam's center until the end of Phase 1 (see Phase 1-3 indicated in \fref{signatureMoveintro}), then a jump to the high-amplitude branch occurs.
From here, the slider turns back towards the beam's center until the end of Phase 2, where a steady state is reached (Phase 3).

\subsection{Resonance condition and phase lag}
The condition $\Omega/\omega=\tilde\omega(s,\hat q)/\omega$ defines a relation between $\hat q$ and $s$ denoted as \emph{\MOD{backbone} curve}, depicted in \fref{sSIM}a.
Using \eref{SHBomtil}, this relation can be given in explicit form:
\ea{
\hat q^{\mathrm{backbone}} = \pm \sqrt{
\frac{4}{3\kappa}\left[
~ \left(1+\mu(s)\right)\left(\frac{\Omega}{\omega}\right)^2 ~ - ~ 1 ~
\right]\fp
}
}
The \sSIM does not intersect with the \MOD{backbone} curve for the given parameters.
In fact, the main or \emph{high-amplitude branch} approaches the \MOD{backbone} curve for higher amplitudes, but never reaches it in \fref{sSIM}a.
The low-amplitude branch has a turning point, and forms an isolated bubble with the unstable intermediate branch.
\MOD{
Interestingly, the bifurcation diagram is almost but not exactly mirror symmetric to $s=0.5$.
This is because the the diagram shows the beam's amplitude at an off-center position ($\hat w(4/7)$), and the mass and inertia added by the slider lead to a slightly asymmetric modal deflection shape for $s\neq 0.5$.
}
\\
An intersection between \sSIM and \MOD{backbone} curve is generally possible for sufficiently low excitation level.
\MOD{
This is shown in \fref{sSIM}b.
In that case, the high-amplitude branch has a resonance peak, and a second turning point occurs, connecting the high-amplitude with the intermediate branch.
The resonance peak appears with a certain distance from the beam's center.
Further, that resonance peak is close to the second turning point (light damping).
Thus, the high-amplitude branch has a rather small basin of attraction near the highest vibration level.
}
In contrast, when the \sSIM has the form as in \fref{sSIM}a, the high-level regime was found to be quite robust, both in numerical and experimental practice.
From this, it is concluded that the self-adaptive system should be driven near but never at resonance; \ie, it is \emph{not desirable to reach the condition of resonance}.
\\
For the analysis of the locomotion along the \sSIM, it is useful to know the phase lag between excitation and response.
\MOD{
The phase lag can be determined from \eref{SHB}:
\ea{
\tan \theta = \frac{2D\frac{\Omega}{\omega}}{(1+\mu)\left(\frac{\Omega}{\omega}\right)^2-1-\frac34\kappa\hat q^2}\fp \label{eq:theta}
}
To estimate the value of $\theta$,} the orders of magnitude of important parameters are considered.
For bookkeeping, a small parameter $\epsilon$ with $10^{-2}\ll \epsilon\ll 1$ is introduced.
In the interesting range, one has $\Omega/\tilde\omega = 1+\epsilon\sigma^*$ with $\sigma^*=\mathcal O(1)$; \ie, $\sigma^*$ is of the order of magnitude of $1$.
Thus, $(\Omega/\tilde\omega)^2 = 1+2\epsilon\sigma^*+\mathcal O(\epsilon^2)$.
At the same time, one has very light damping $D=0.1~\% = \mathcal O(\epsilon^2)$ (\tref{params}).
With this, \MOD{the numerator on the right-hand side of \eref{theta} is much smaller than the denominator.}
This means that one has either $\theta \lessapprox 0$, or $\theta\gtrapprox-\pi$.
The further the slider is near the clamping, the higher the natural frequency.
Consequently, the high-amplitude branch, which is closer to the clamping than the \MOD{backbone} curve, has $\tilde\omega>\Omega$, \ie, $\sigma^*<0$, so that $\theta \lessapprox 0$ can be followed from \MOD{\eref{theta}}.
This means that the displacement response is approximately in phase with the base displacement on the high-amplitude branch.
In contrast, the branch closer to the beam's center has $\tilde\omega<\Omega$, \ie, $\sigma^*>0$, so that $\theta\gtrapprox-\pi$.
This means that the displacement response is approximately in anti-phase with the base displacement on the low-amplitude branch.
%

\subsection{Validation of the approximated \sSIM\label{sec:sSIMvalidation}}
In the above described theory, the modal deflection shape of the beam with ideal boundary conditions and without the slider was used in the modal truncation.
The effect of the non-ideal boundary conditions and the attached slider on the modal deflection shape is illustrated in \fref{modalDeflectionShape}.
While the non-ideal clamping (finite rotational stiffness) has a minor effect only, the influence of the attached slider mass and rotational inertia is moderate.
The influence of the different mode shape on the \sSIM is illustrated in \fref{sSIM}c.
While the qualitative behavior is the same, the \sSIM appears to be shifted/stretched towards the clamping compared to the simplified theory.
\\
In reality, the slider is neither permanently attached nor permanently disconnected from the beam, and the impacts add dissipation.
To account for the dynamic contact interactions between slider and beam, the Pseudo-Constrained Slider (PCS) model is simulated.
\MOD{
This model was proposed in \cite{Muller.2023} and is briefly summarized in the following to make the article self-contained.
In the PCS model, the slider is let horizontally free, but the prescribed (relative) slider position is always fed to the algorithm that evaluates the contact kinematics.
}
This can be interpreted as moving the beam horizontally according to the slider's motion.
This PCS model is to be distinguished from the model with free slider (FS model), where the slider can re-arrange along the beam.
For the numerical simulations throughout this work, the procedure described in \cite{Muller.2023} is used, where unilateral interactions in the contact normal direction and Coulomb dry friction in the tangential direction are modeled at the four possible contact points (\fref{SAS}b).
\\
\MOD{
The results of the PCS model are depicted in \fref{sSIM}d.
The green curve is obtained by step-wise increasing $s$ in the model from $s=0$ to $s=0.5$.
In each step, the simulation is run until a sufficiently steady state has been reached.
Of a sufficiently long recording of the steady state, an amplitude measure for the beam vibration is then extracted using the Hilbert transform (window size $1~\mrm{s}$).
The current values of the coordinates and velocities at the end of one step are adopted as initial values for the next step.
The yellow curve is obtained in an analogous way, only by backward stepping from $s=0.5$ to $s=0$.
Homogeneous initial conditions are provided for the very first step.
Because the procedure relies on forward time stepping, only asymptotically stable states can be obtained.
This explains why the yellow curve exhibits a jump where a turning point would be expected according to the analytical approximation.
Indeed, the results of the PCS model are quite close to the analytical approximation.
}

\subsection{Resemblance to host structure with impact absorber; strongly modulated vs. almost periodic response}
The setup, consisting of a vibrating flexible structure, carrying a mass with free play, resembles that of a host structure with an impact vibration absorber.
Impact absorbers are typically placed into a cavity so that the impacts occur inside the host structure.
In contrast, the impacts occur outside (on the surface) of the host structure (beam) in the case of the self-adaptive system.
However, this is merely a matter of technical implementation and has no effect on the system dyamics.
An important difference, besides the obvious ability of the self-adaptive system to re-arrange on the super-slow time scale, is that the mass of an absorber is usually much smaller (typically $\ll 10\%$) than that of the host structure, whereas the slider is actually more heavy here (mass ratio $m/(\rho AL)\approx 300\%$).
Another important difference is that the slider has two contact points on each side instead of just one, which gives rise to much more complicated contact sequences if relative rotation (\emph{rocking}) occurs between slider and beam.
\\
A well-known result from the theory of impact absorbers, or more generally Nonlinear Energy Sinks, in the presence of periodic forcing, is the existence of two regimes, \emph{strongly modulated} and \emph{almost periodic response} \cite{Vakakis.2009}.
For sufficiently high vibration levels, the vibrations are almost periodic.
In a range of lower vibration levels, on the other hand, strongly modulated vibrations occur.
This is exactly as observed in the self-adaptive system \cite{Muller.2023}, see also \sref{locovalidation}.
\\
For the case of small mass ratios between an absorber and a single-degree-of-freedom host structure, a closed-form expression was derived by Gendelman \cite{Gendelman.2015} for the vibration level needed to permit almost periodic behavior.
In that case, the vibrations of the host structure are in good approximation harmonic, and the absorber is assumed to follow a triangle wave function of the same period, undergoing two symmetric impacts per period.
For a given phase between harmonic movement of the host structure's walls and the triangle wave, one obtains a certain ratio between post- and pre-impact relative velocity, which is related to the restitution coefficient $r$.
For a phase of zero, for instance, post- and pre-impact velocity are equal (restitution coefficient $r=1$).
For a given restitution coefficient $r$ and clearance $g$, one obtains a minimum amplitude of the wall movement, $\hat w(s)$, below which the triangle wave is kinematically inadmissible.
Below this threshold, (almost) periodic behavior is not possible and instead the response becomes strongly modulated.
The condition takes the form:
\ea{
\hat w(s) \geq g\frac{\overline\rho}{\sqrt{1+\overline\rho^2}} \fk \quad \overline\rho = \frac2\pi \frac{1-r}{1+r} \fp \label{eq:rho}
}
The threshold increases monotonically with decreasing $r$.
For $r\to 1$, one has $\overline\rho\to 0$, so that the condition reads $\hat w(s)\geq 0$; \ie, strongly modulated behavior is not expected.
For $r=0.5$, which was used in the simulation of the self-adaptive system \cite{Muller.2023}, one has $\overline\rho = 2/(3\pi)$ so that $\hat w(s) \geq 0.21g$.
Using some $r<0.5$ in \eref{rho} seems justified because the theory assumes a single-degree-of-freedom host structure, whereas many modes were retained in the simulation, and the transfer of energy to the higher-frequency modes could be mimicked by a smaller restitution coefficient.
For the limit of $r\to 0$, one has $\overline\rho\to 2/\pi$, yielding $\hat w(s)\geq 0.54g$.
In the test rig of the self-adaptive system, the clearance is $g=(R-h)/2=h(R/h-1)/2=0.025h$ with $h/L=0.0071$ (\tref{params}).
For $r=0.5$ ($r=0$), one obtains the threshold $\hat w(s)/L\geq 4\cdot 10^{-5}$ ($\hat w(s)/L\geq 1\cdot 10^{-4}$).
The actually encountered strongly modulated response on the low-amplitude branch occurs at a vibration level that is about one order of magnitude larger.
Apparently, the important differences (large instead of small mass ratio; two contacts per side instead of one), are responsible for this discrepancy.
Still, it is postulated that the qualitative observations (low vibrations: strongly modulated response; high vibrations: almost periodic response), are due to kinematic constraints analogous to Gendelman's theory \cite{Gendelman.2015}.

\section{Locomotion: Theory\label{sec:locotheory}}
The main focus of the present work is the locomotion of the slider along the beam.
In this section, hypotheses on the dominant forms of locomotion are formulated.
These are validated in the subsequent \sref{locovalidation}.
The system has a considerable number of geometric and material parameters, and can be exposed to a wide range of excitation and initial conditions.
The point of departure for the subsequent analysis is the recently validated model with the parameters listed in \tref{params}.
The analysis is limited to the three different phases encountered during the \emph{signature move}, as indicated in \fref{locomotion_temp_separable}.
Recall that $s=x^{(Q)}/L$, and $\Delta s$ corresponds to the horizontal slider transport per excitation period.
During each phase, a certain reference situation is defined, denoted as \emph{Case} 1-3, representative of \emph{Phase} 1-3, respectively, and indicated as blue squares in \fref{locomotion_temp_separable}.
The locomotion observed in those situations was found to be representative also of the active locomotion encountered under all other simulated excitation and initial conditions.
The excitation parameters and important response quantities are listed in \tref{signaturemoveparams} in dimensionless form.
Those values are used throughout \sref{locotheory}, and they were obtained by the PCS model explained in \ssref{sSIMvalidation}.
Very similar values would be obtained with the analytical approximation, but the values of the more accurate PCS model are preferred here to avoid an additional error source.
\onefig{1}
{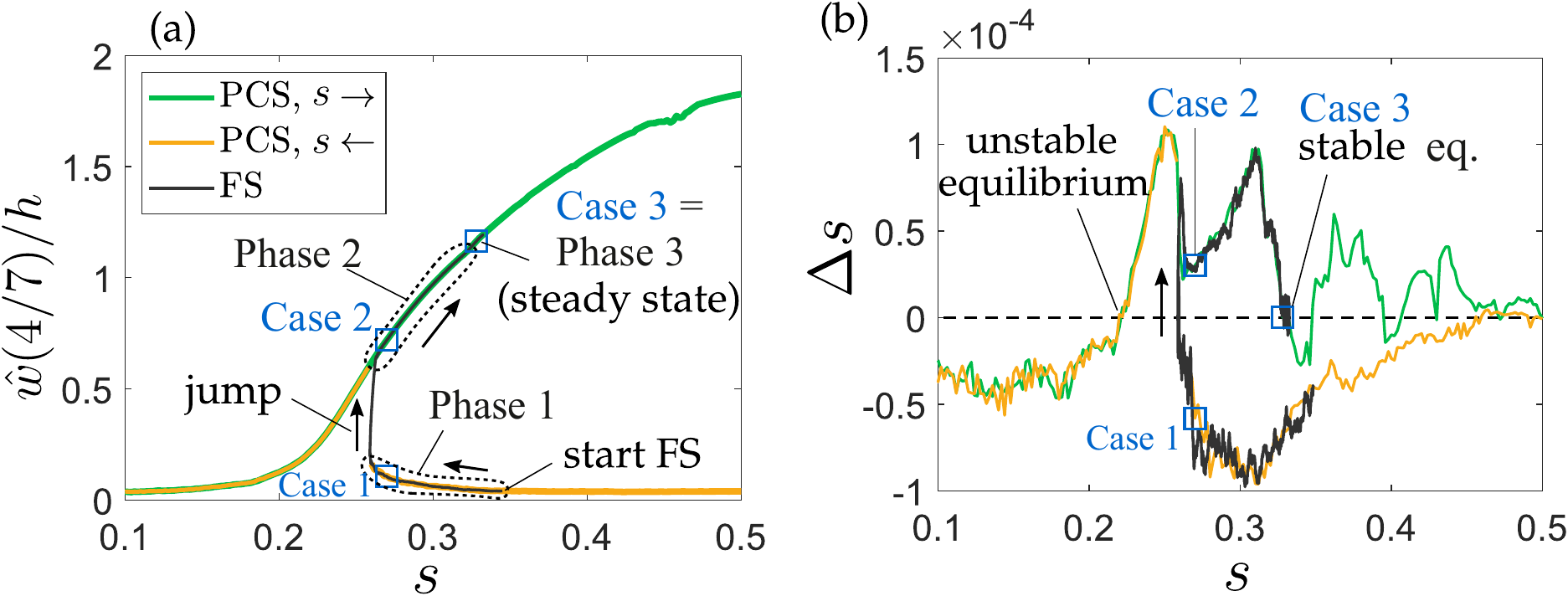}
{Comparison of super-slow dynamics resulting from PCS and FS models, blue squares indicate the reference situations (Case 1-3), representative of Phase 1-3, respectively: (a) beam's vibration level versus horizontal slider position, (b) horizontal slider transport per period.
PCS: Pseudo-Constrained slider; FS: Free Slider
}
{locomotion_temp_separable}
\tab[htb]{lllll}{
Quantity & Symbol & Value (Case 1) & Value (Case 2) & Value (Case 3) \\
\hline
excitation level & $\hat w^{(0)}/L$ & $1.65\cdot 10^{-4}$ & $1.65\cdot 10^{-4}$ & $1.65\cdot 10^{-4}$ \\
excitation frequency & $\Omega/\omega$ & $0.477$ & $0.477$ & $0.477$ \\
slider location & $s$ & $0.27$ & $0.27$ & $0.33$ \\
displacement magnification & $\hat{w}(s)/\hat{w}^{(0)}$ & $6.50$ & $30.4$ & $54.2$ \\
at slider location &&&&\\
displacement amplitude & $\hat w(s)/L$ & $0.0011$ & $0.0050$ & $0.0090$ \\
at slider location &&&&\\
clearance ratio & $2\hat w(s)/(R-h)$ & $6$ & $28$ & $50$ \\
\hline
slope amplitude & $\hat w_\xi(s)/L$ & $0.0053$ & $0.0246$ & $0.0296$ \\
at slider location &&&&\\
curvature amplitude & $\hat w_{\xi\xi}(s)/L$ & $0.0086$ & $0.0400$ & $0.1171$ \\
at slider location &&&&\\
ratio displacement ampl. & $\phi(s^{(\mrm r)})/\phi(s^{(\mrm l)})$ & $1.43$ & $1.43$ & $1.27$ \\
right/left contact &&&&\\
displacement amplitude & $\hat q$ & $0.0018$ & $0.0083$ & $0.0117$\\
at beam center &&&&
}{Dimensionless dynamic variables during the signature move.
$s^{(\mrm r)} = s+B/(2L)$, $s^{(\mrm l)} = s-B/(2L)$.
All quantities from the top down to the clearance ratio were obtained from the PCS model and the geometric parameters listed in \tref{params}.
The remaining parameters are estimated based on the analytical approximation derived in \sref{sSIM}.
}{signaturemoveparams}
\\
In the reference situations, the elastic displacement of the beam is one to two orders of magnitude larger than the imposed base motion, as expected near resonance of a lightly-damped structure.
Thus, the base motion does not play an important role in the locomotion there.
The elastic displacement is also much larger than the clearance.
Actually, the ratios are similar, which implies that the clearance is in the order of magnitude of the amplitude $\hat w^{(0)}$ of the base displacement (\tref{signaturemoveparams}).
The beam's elastic displacement at the slider location is two to three orders of magnitude smaller than its length.
With $h/L\approx 7.1\cdot 10^{-3}$ (\tref{params}), the displacement amplitude at the slider location is an order of magnitude smaller than the thickness in Case 1, while it is of the same order of magnitude in Case 2-3.
In all cases, the slope is rather small, leading to an angle of less (or even much less) than $2^\circ$.
Thus, many kinematic expressions can be simplified assuming small rotations.

\subsection{Can gravity impede locomotion?\label{sec:inactivity}}
Consistent with \sref{sSIM}, it is assumed in large parts of \sref{locotheory} that the beam displacement remains in good approximation harmonic.
If the transverse acceleration of the beam is smaller than the gravity acceleration, the upper contacts are expected to be closed permanently and sticking.
In that case, no vibration-induced locomotion is expected (\emph{inactivity}).
The transverse acceleration is composed of the base acceleration $w^{(0)}_{tt}$ and the elastic acceleration $w_{tt}(s)$.
For the considered excitation conditions, the base acceleration already exceeds the gravity acceleration ($\hat w_{tt}^{(0)}=14~\mrm{m}/\mrm{s}^2>9.81~\mrm{m}/\mrm{s}^2$).
As established in \sref{sSIM}, $w$ and $w^{(0)}$ are in phase along the high-amplitude branch, so that the beam's total transverse acceleration, $|w^{(0)}_{tt}+w_{tt}|\geq |w^{(0)}_{tt}|$.
Thus, no regime of inactive locomotion is expected along the high-amplitude branch.
On the low-amplitude branch, $w$ and $w^{(0)}$ are in anti-phase, so that destructive interference is possible.
Indeed, for the range $0.33<s<0.5$, the elastic displacement is only about a factor two larger than the base displacement, so that $\hat w_{tt}(s) \approx 2 \hat w^{(0)}_{tt}$.
Thus, very slow or no locomotion is expected in that part of the low-amplitude branch.
For $s<0.33$, the vibration level increases quickly so that the elastic acceleration dominates, $\hat w_{tt}(s) \gg \hat w^{(0)}_{tt}$, so that locomotion is feasible in that part of the low-amplitude branch. 

\subsection{Expected contact sequences\label{sec:contactSequence}}
For the description of contact sequences, it is useful to label the four contacts ($P_1$ to $P_4$) of the slider, and assign markers (\fref{allocation_cts_pts}).
We shall exclude the case where gravity impedes locomotion, and consequently sufficiently large vibrations are assumed so that the upper contacts are not permanently closed.
In the half period with positive displacement, $w(s)>0$, the acceleration is negative, $w_{\tau\tau}(s)<0$ (recall that harmonic vibration is assumed).
As the contacts can only transmit compressive forces in the normal direction, the upper contacts are expected to open (if they were closed initially) and the lower contacts are expected to close near the beginning of that half period, and remain closed until the end of that half period.
The exact opposite holds for the other half period ($w(s)<0$, $w_{\tau\tau}(s)>0$, upper contacts tend to be closed).
Near $w\approx 0$, a switch between lower/upper contacts is expected.
Assuming that a free-flight phase occurs here, the duration between the opening of one set of contacts and the closing of the other depends on the clearance and the slider's velocity (equal to the beam's velocity when the contacts open).
The smaller the clearance is compared to the transverse displacement level, the shorter this duration.
\onefig{.265}
{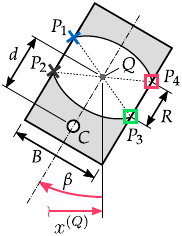}
{Schematic of the slider showing the assignment of the markers to the contact points.}
{allocation_cts_pts}
\\
The above discussion holds strictly only as long as there is no relative rotation between slider and beam.
However, the beam has a certain slope at the slider location, and, because of the finite slider width $B$, there is a considerable difference between the displacement level of left and right contacts.
More specifically, the contacts closer to the beam's center (location of anti-node) are exposed to an approximately $30-40~\%$ larger displacement amplitude than the contacts further away from the beam's center (\tref{signaturemoveparams}).
This has the potential to induce relative rotations (\emph{rocking}).
The rocking leads to situations where only a single instead of both upper/lower contacts are closed.
For sufficiently large relative rotation, contacts may also close diagonally (upper-left/$P_1$ and lower-right/$P_3$ closed, or upper-right/$P_4$ and lower-left/$P_2$ closed).
This condition marks the maximum relative rotation and will be referred to as \emph{pitch limit}.
Reaching the pitch limit is crucial for an important form of locomotion which will be referred to as \emph{pitching cycle}.
This form of locomotion works fine even when the closed contacts are sticking; \ie, sliding is not needed.
In contrast, the other forms of locomotion heavily rely on sliding; \ie, it is crucial to overcome static dry friction.
On the other hand, those forms of locomotion work fine when the pitch limit is not reached.
It will be shown that the horizontal acceleration that causes the sliding is composed of two important terms, the \emph{slope-induced acceleration} and the \emph{rocking-induced acceleration}.
The former is already present if no relative rotation between beam and slider is possible, whereas the latter relies exactly on this relative rotation.
In the following, first, the pitching cycle is presented, and a closed form expression is derived for the horizontal slider transport per cycle under the assumption of sticking contacts.
Subsequently,
the sliding due to slope- and rocking-induced acceleration is analyzed.

\subsection{The pitching cycle\label{sec:pitchingCycle}}
\onefig{.7}
{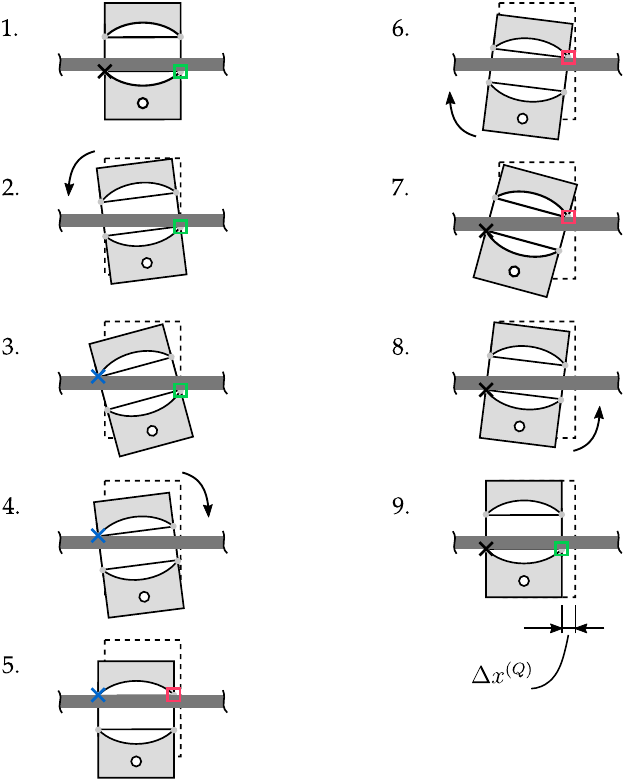}
{Sequences of the idealized pitching-locomotion process.
}
{visualization_idealized}
%
The pitching cycle is schematically illustrated in \fref{visualization_idealized}, in the coordinate system translating and rotating with the beam.
The shown slider motion is periodic in the sense that the slider's transverse displacement and rotation, respectively relative to the beam, repeat each cycle, whereas the horizontal position of the contact center ($x^{(Q)}$) varies monotonically.
The process consists of rotations around one contact, separated by transitions where two contacts are closed at the same time.
All combinations of active contacts that are kinematically possible are part of the process.
This includes the cases of minimum and maximum rotation (relative to the beam); \ie, the pitch limit is reached (states 3 an 7 in \fref{visualization_idealized}).
One can directly see that the depicted kinematic process causes transport of the slider to the \textit{left} side.
In the following, the distance $\Delta s^{\mrm{pitch}} = \Delta x^{(Q)}/L$ that the slider advances per cycle, is approximated.
To this end, the pitch limit must be expressed first.
The interpretation of the pitching cycle from a kinetic point of view, \ie, the explanation why the cycle evolves in this particular way, is analyzed in \sref{locovalidation}.
\\
Using the theory derived in \sref{sSIM}, one can easily verify that the beam is in very good approximation straight between left and right contacts.
More specifically, the fraction of the transverse displacement amplitude, which stems from the non-straight deformation, is small compared to the clearance,
$(\phi(s^{(\mrm r)})/\phi(s^{(\mrm l)})-1)\hat w(s)/L-\hat w_\xi(s)/L~B/L \ll 0.05h/L$, where $s^{(\mrm r)}$ and $s^{(\mrm l)}$ are defined in the caption of \tref{signaturemoveparams}.
Thus, rigid body kinematics can be used for deriving the pitching cycle, and the beam is represented as straight in \fref{visualization_idealized}.
The pitch limit, $\hat{\beta}_\mrm{rel}$, between slider and beam, hence depends purely on the geometric parameters $B$, $R$, $h$:
\e{
\hat{\beta}_\mrm{rel} = \mrm{atan} \left( \frac{R}{B} \right) - \MOD{\mrm{arcsin}} \left(\frac{h}{\sqrt{R^2+B^2}}\right)\fp
}{beta_r_lim}
For the geometric parameters given in \tref{params}, one obtains a pitch limit of $\hat{\beta}_\mrm{r}\approx 0.29^\circ$.
\tab[htb]{llllll}{
Quantity & Symbol & Value $\Pone$ & Value $\Ptwo$ & Value $\Pthree$ & Value $\Pfour$ \\
\hline
pitch limit & $\hat{\beta}_{\mrm{rel}}$ & $5\cdot 10^{-3}$ & $5\cdot 10^{-3}$ & $5\cdot 10^{-3}$ & $5\cdot 10^{-3}$ \\
vertical indicator & $\zeta_{z}$ & $+1$ & $-1$ & $-1$ & $+1$ \\
horizontal indicator & $\zeta_{x}$ & $-1$ & $-1$ & $+1$ & $+1$ \\
relative pendulum length & $\ell/L$ & $0.0487$ & $0.0439$ & $0.0439$ & $0.0487$\\
pendulum angle & $\sigma$ & $-0.8243$ & $-0.9501$ & $0.9501$ & $0.8243$ \\
$\sin(2\sigma)/2$ & $-$ & $-0.4985$ & $-0.4731$ & $0.4731$ & $0.4985$ \\
relative rotary inertia & $m\ell^2/(J^{(C)}+m\ell^2)$ & $0.4049$ & $0.3565$ & $0.3565$ & $0.4049$
}{Dimensionless geometric parameters related to locomotion. Angles are actually in $\mathrm{rad}$.}{locoparams}
\\
Assuming that at least one contact is always closed (no free flight), and that closed contacts are always sticking (no sliding), the horizontal transport per cycle follows from a purely kinematic analysis.
Due to the symmetry, the transport can be broken down to four identical portions, and reads as
\ea{
|\Delta s^{\mrm{pitch}}| &=& \frac{1}{L} 4 \frac12\Big( R \sin \hat{\beta}_\mrm{rel} - B \left( 1 - \cos \hat{\beta}_\mrm{rel} \right) \Big)\fk \label{eq:DeltaXQfull}\\
& \approx & 2\frac{R}{L}\hat{\beta}_{\mrm{rel}}\fp \label{eq:DeltaXQ}
}
%
In the last step, it was exploited that $0<\hat{\beta}_{\mrm{rel}}\ll 1$, so that $\sin \hat{\beta}_\mrm{rel}\approx \hat{\beta}_{\mrm{rel}}$ and $\cos \hat{\beta}_\mrm{rel}\approx 1$.
For the parameters given in \tref{params}, $|\Delta s^{\mrm{pitch}}|=7.5\cdot 10^{-5}$.
To illustrate this number: It would take around 100 cycles to travel one beam thickness $h$ further, or 1000 cycles to travel one slider width $B$ further.
It will be shown that the pitching cycle is the main driver of the locomotion along the low-amplitude branch.
Thus, it is not surprising that $\Delta s$ in Phase 1 is in the predicted order of magnitude (\fref{locomotion_temp_separable}b).
The sign of $|\Delta s^{\mrm{pitch}}|$, \ie, the direction of transport induced by the pitching cycle is explained in \sref{locovalidation}.
%
\\
It is worth noting that similar pitching locomotion has already been observed in other non-smooth systems with rotational degrees of freedom, such as the woodpecker toy or a certain tumbling toy \cite{Leine.2003}.
The pitching cycle can be interpreted as swing motions of a pendulum where the hinge point changes at the pitch limits.
In that sense, there is also some similarity to the brachiation of primates, which also can be modeled as pendular locomotion \cite{Preuschoft.1985,Preuschoft.2022}.
The abstraction to a pendulum also plays a crucial role in the subsequently analyzed forms of locomotion.

\subsection{Sliding due to slope- and rocking-induced acceleration\label{sec:rotindsl}} 
Besides the pitching cycle, sliding-type locomotion is also possible, and this is mainly caused by two important acceleration terms.
The rocking-induced acceleration requires the possibility of relative rotation between slider and beam.
Thus, it is only active if a single contact is closed.
The slope-induced acceleration occurs not only if a single contact is closed, but also when two contacts are closed (upper or lower or diagonal).
The slope-induced acceleration takes the form postulated by Thomsen \cite{Thomsen.1996}, while its exact value depends on the contact situation, as described later.
In the following, both acceleration terms are studied by considering a single contact as closed.
The slider then hinges on this contact, and acts as a pendulum attached to a moving base, as illustrated in \fref{pendulum}.
The slider is free to rotate and move horizontally along the beam, while the motion of the hinge point is constrained to the beam in the transverse direction.
For simplicity, the hinge point is permitted to slide without friction in the following.
The discussion whether or not the tangential accelerations are sufficient to overcome static dry friction is postponed to the end of this subsection.
\onefig{.42}
{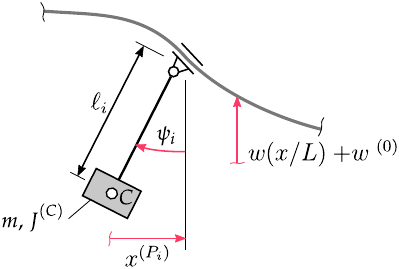}
{Pendulum hinged to a point that is free to slide along a vibrating beam.
}
{pendulum}

\subsubsection{Equation of motion of the sliding pendulum}
The geometric relations differ slightly from contact to contact.
More specifically, the pendulum length is
\e{\ell_i=\sqrt{\left(\frac{B}{2}\right)^2+\left(d+\zeta_{z,i}\frac{R}{2}\right)^2}\fk
}{length_pendulum}
where $\zeta_{z,i}=+1$ for the upper and $\zeta_{z,i}=-1$ for the lower contacts.
The rotation angle is defined as
\ea{
\psi_i(t) = \sigma_i + \beta(t)\fk \label{eq:psi}
}
where $\beta$ measures the rotation of the slider in the inertial frame of reference (\fref{SAS}b), and $\sigma_i= \zeta_{x,i} \ \MOD{\mrm{arcsin}}\left(B/(2\ell_i)\right)$, with $\zeta_{x,i}=+1$ for the right and $\zeta_{x,i}=-1$ for the left contacts.
$\psi_i$ is the angle between the connection line from hinge point to center of mass and the vertical direction (\fref{pendulum}).
The design values of the pendulum length $\ell_i$ and angle $\sigma_i$ are listed in \tref{locoparams} for the four different contacts.
In the following, the index $i$ is omitted for readability.
\\
The equation of motion of the sliding pendulum is derived using the Lagrange method.
As already discussed, the beam's transverse acceleration is much larger than the gravity acceleration, so that the latter is neglected in the following.
Thus, the pendulum's potential energy is constant.\footnote{The term \emph{pendulum} may seem awkward when gravity is neglected. In the present case, the acceleration imposed by the base motion plays a similar role as gravity, which is why we still find the term \emph{pendulum} appropriate.}
Also, there are no causes of non-conservative forces, when considering only the slider without friction.
As a consequence, the kinetic energy is conserved, which reads as
\ea{
E_{\mrm{kin}} &=& \frac12 J^{(C)} \psi^2_{tt} + \frac12 m\left( \left(x_t^{(C)}\right)^2 + \left(y_t^{(C)}\right)^2 \right) \fk \label{eq:Ekin} \\
x^{(C)} &=& pL - \ell\sin\psi \fk \label{eq:xC} \\
y^{(C)} &=& w^{(0)} + w^{\mrm P} - \ell\cos\psi \fk \label{eq:yC} \\
x_t^{(C)} &=& x_t^{\mrm P} - \ell \psi_t \cos\psi \fk \label{eq:xCdot} \\
y_t^{(C)} &=& w_t^{(0)} + w^{\mrm P}_t + w_{x}^{\mrm P}x_t^{\mrm P} + \ell \psi_t \sin\psi\fp \label{eq:yCdot}
}
Herein, $p=x^{(P)}/L$ is the normalized horizontal position of the contact point.
Note that the beam is thin and its rotation is small, in the sense that $w_xh/(2L)\ll 1$.
Consequently, the corresponding transverse displacement of the beam is simply $w^{\mrm P} = w(p)$; \ie, the actual contact point on the beam's lower/upper surface has approximately the same displacement as the corresponding point on the beam's neutral axis.
The analog holds for the derivatives; \ie, $w^{\mrm P}_t$ and $w^{\mrm P}_x$ are also evaluated at $p$.
\\
Using $p$ and $\psi$ as generalized coordinates, one obtains the equations of motion:
\ea{
& \psi_{\tau\tau} +  \frac{m \ell^2}{\left( J^{(C)} + m \ell^{2} \right)} \frac{L}{\ell} \Big( \frac{w_{\xi}^{\mrm P}}{L} p_{\tau\tau} \sin\psi  + \frac{w_{\xi\xi}^{\mrm P}}{L} p_\tau^{2} \sin\psi  \nonumber\\
& \quad\quad\quad\quad + 2 \frac{w_{\xi\tau}^{\mrm P}}{L} p_\tau \sin\psi   - p_{\tau\tau} \cos\psi  +  \frac{w_{\tau\tau}^{(0)}+w_{\tau\tau}^{\mrm P}}{L} \sin\psi  \Big) = 0\fk \label{eq:EOM_psi_pendulum}
\\
& p_{\tau\tau} + \frac{\ell}{L} {\psi}^{2}_\tau \sin\psi\left(1+\frac{w_\xi^{\mrm P}}{L}\cot\psi\right) - \frac{\ell}{L} {\psi}_{\tau\tau} \cos\psi\left(1-\frac{w_\xi^{\mrm P}}{L}\tan\psi\right) \nonumber\\
& \quad\quad\quad\quad  + \big( \frac{w_{\xi\xi}^{\mrm P}}{L} p_\tau^2 + 2 \frac{{w}_{\xi\tau}^{\mrm P}}{L} p_\tau + \frac{w_\xi^{\mrm P}}{L} p_{\tau\tau} + \frac{w_{\tau\tau}^{(0)}+w_{\tau\tau}^{\mrm P}}{L} \big)\frac{w_\xi^{\mrm P}}{L}
 = 0\fp
\label{eq:EOM_p_pendulum}
}
These equations are expressed in dimensionless form.
Note that all terms in these equations are accelerations, so that all derivatives with respect to $t$ can be simply replaced by those with respect to any time variable, and the equations will still be valid.
The vibration drives the locomotion, which occurs on the time scale of the excitation.
It is thus convenient to \emph{redefine the normalized time variable} as $\tau=\Omega t$ in this section ($\tau=\omega t$ was used in \sref{sSIM}).

\subsubsection{Identification of dominant acceleration terms}
To understand what really moves the slider along the beam, the most important terms in \erefs{EOM_psi_pendulum}-\erefo{EOM_p_pendulum} need to be identified.
As noted before, the base displacement is one or even two orders of magnitude smaller than the elastic displacement in the reference cases (\tref{signaturemoveparams}), and is thus neglected in the following.
Second, one may notice that the slope $\frac{w_\xi^{\mrm P}}{L}$ is small:
Using the approximation $s\approx p$, one can infer from \tref{signaturemoveparams} an approximate upper bound, $\frac{|w_\xi^{\mrm P}|}{L}<0.03$.
Third, one may notice that the rotation $\beta$ is small in the sense that $|\beta| \ll |\sigma|$:
An upper bound is $|\beta|\leq \frac{\hat w_\xi}{L} + |\hat{\beta}_{\mathrm{rel}}| < 0.035$, where $|\hat{\beta}_{\mathrm{rel}}|=0.005$, while $|\sigma|>0.82$ (\tref{locoparams}).
Thus one can use the approximation $\psi=\sigma+\beta\approx\sigma$ in all arguments of the trigonometric functions in \erefs{EOM_psi_pendulum}-\erefo{EOM_p_pendulum}. 
However, since $\sigma$ is time-constant, $\beta$ cannot be neglected in the time derivatives of $\psi$; \ie, $\psi_\tau=\beta_\tau$, $\psi_{\tau\tau}=\beta_{\tau\tau}$.
Further, assuming that $\beta$ oscillates with dominant frequency $\Omega$, then $\psi_\tau$ and $\psi_{\tau\tau}$ will be of the same order of magnitude as $|\beta|$.
Consequently, $|\psi_\tau^2|\ll|\psi_{\tau\tau}|$.
With the values of $\sigma$ listed in \tref{locoparams}, $\sin\psi$ and $\cos\psi$ are of similar order of magnitude, and both $\tan\sigma$ and $\cot\sigma$ are of order of magnitude one.
From the above, it follows that the terms in the parenthesis in the top of \eref{EOM_p_pendulum} are approximately unity ($|\frac{w_\xi^{\mrm P}}{L}\cot\psi|\ll1$, $|\frac{w_\xi^{\mrm P}}{L}\tan\psi|\ll 1$), and that the second term in \eref{EOM_p_pendulum} is negligible.
\\
In accordance with the current state of knowledge, $s$ and thus $p$ change super-slowly with time.
$p$ will turn out to consist of a small oscillatory component with dominant frequency $\Omega$, and an even smaller monotonic component.
Consequently, $p_\tau$ and $p_{\tau\tau}$ are of equal order of magnitude.
The precise order of magnitude can be estimated from \eref{EOM_p_pendulum} as the order of magnitude of the term $\psi_{\tau\tau}\frac{\ell}{L}$, plus that of the term $\frac{w_{\tau\tau}}{L}\frac{w_{\xi}}{L}$.
The former is $10^{-3}$, whereas the latter is always $<3\cdot 10^{-4}$ according to the parameters listed in \tref{signaturemoveparams} (a detailed quantitative derivation follows later).
With the given values, $p_\tau$ and $p_{\tau\tau}$ are of the order of magnitude of $10^{-3}$.
It can be followed that all terms in \eref{EOM_psi_pendulum} with $p_\tau$ or $p_{\tau\tau}$ are negligible.
The same holds for the terms with $p_\tau$ or $p_{\tau\tau}$ in the last parenthesis of \eref{EOM_p_pendulum}.
All above simplifications together yield the approximation of the rotational ($\psi_{\tau\tau}$) and the horizontal ($p_{\tau\tau}$) acceleration:
\ea{
\psi_{\tau\tau} &\approx& -\frac{w_{\tau\tau}^{\mrm P}}{L} \frac{L}{\ell} \frac{m\ell^2}{J^{(C)}+m\ell^2}\sin\sigma \fk \label{eq:psiddotapprx} \\
p_{\tau\tau} &\approx& \underbrace{-\frac{w_{\tau\tau}^{\mrm P}}L \frac{w_{\xi}^{\mrm P}}L}_{p_{\tau\tau}^{\mrm{{slope}}}} \underbrace{-\frac{w_{\tau\tau}^{\mrm P}}{L}\frac{m\ell^2}{J^{(C)}+m\ell^2}\frac{\sin\left(2\sigma\right)}{2}}_{p_{\tau\tau}^{\mrm{rock}}} \fp \label{eq:sloperot}
}
The horizontal acceleration is composed of two terms, the \emph{slope-induced acceleration} and the \emph{rocking-induced acceleration}.
The former is identical to the term derived in \cite{Thomsen.1996,Miranda.1998} for a point mass on a vibrating beam/string\footnote{More precisely, the terms are identical for $p=s$, which is the case for a point mass / slider with negligible width $B\ll L$.}.
This term can be interpreted as the beam's transverse acceleration projected onto the rotated beam axis.
Consequently, this term is already present if no clearance was allowed between beam and slider / the slider's rotation was constrained to that of the beam.
In contrast, the term $p_{\tau\tau}^{\mrm{rock}}$ requires that the slider's rotation is a degree of freedom; \ie, relative rotation between beam and slider is not constrained.
Further, the line connecting the hinge point of the pendulum (contact point) and the center of mass must not be vertical nor horizontal:
For $\sigma=0$ (vertical), there is no rotational acceleration.
For $\sigma=\pi/2$ (horizontal), the rocking-induced acceleration points in the vertical direction (and hence does not affect the slider's horizontal movement along the beam).
\\
As derived in \sref{sSIM}, the beam's acceleration $w_{\tau\tau}$ is in anti-phase with its displacement $w$.
Due to symmetry, it is sufficient to consider the left half of the beam only.
In that section, the slope $w_\xi$ has the same sign as the displacement $w$.
Thus, the product $w_{\tau\tau}w_\xi<0$ is strictly negative; the slope-induced acceleration always points towards the beam's center.
\\
In contrast to the slope-induced acceleration, the direction of the rocking-induced acceleration depends on the contact point.
The right contacts have $0<\sigma<\pi/2$, while the left have $-\pi/2<\sigma<0$ (\tref{locoparams}).
Thus, $\sin(2\sigma)$ is positive (negative) for the right (left) contacts.
As discussed in \ssref{contactSequence}, the lower contacts tend to be closed for $w>0$ where $w_{\tau\tau}<0$, and the upper contacts tend to be closed for $w<0$ where $w_{\tau\tau}>0$.
This leads to the following signs of the rocking-induced acceleration:
\begin{itemize}
\item[--] $\Pone$: \ $p_{\tau\tau}^{\mrm{rock}} > 0$ (right slip),
\item[--] $\Ptwo$: \ $p_{\tau\tau}^{\mrm{rock}} < 0$ (left slip),
\item[--] $\Pthree$: \ $p_{\tau\tau}^{\mrm{rock}} > 0$ (right slip), and
\item[--] $\Pfour$: \ $p_{\tau\tau}^{\mrm{rock}} < 0$ (left slip).
\end{itemize}
In parentheses, the expected sliding direction is expressed.
As can be inferred from \erefs{psiddotapprx}-\erefo{sloperot}, the rocking-induced acceleration $p_{\tau\tau}^{\mrm{rock}}$ has the same sign as the rotational acceleration $\psi_{\tau\tau}$.
It can be verified that the sign of $\psi_{\tau\tau}$ is such that the slider tends to rotate so that the other contact on the same (upper or lower) side closes.
Thus, the sliding-pendulum effect decelerates the rotation towards the pitch limit.

\subsubsection{Estimation of the horizontal transport per cycle due to slope- and rocking-induced acceleration}
In the following, it is shown that the rocking-induced acceleration is typically much larger than the slope-induced acceleration.
However, the contributions of the individual contacts to the rocking-induced acceleration partially cancel, so that the slope-induced acceleration still plays an important role.
Next, the order of magnitude of the horizontal acceleration terms is estimated.
Subsequently, the net contribution, averaged over all contacts, to the horizontal transport per cycle is estimated.
\\
Recall that it is assumed that the beam vibrates harmonically in its fundamental mode.
Further, it was established that the lower contacts tend to be closed during the half period with $w>0$, and the upper contacts tend to be closed during the half period with $w<0$.
The slope-induced acceleration will be active whenever one or two contacts are closed.
Its exact value depends on $p$; $p$ enters as argument of $w$ and its derivatives in \eref{sloperot}.
When one of the left contacts is closed, one has $p=s^{(\mrm l)}$, when one of the right contacts is closed, one has $p=s^{(\mrm r)}$, whereas when both upper or lower contacts are closed, or a pair of diagonal contacts is closed, one should use $p=s$.
Provided that the contact sequence is in average symmetric, the approximation $p\approx s$ is reasonable even in situations where just one contact is closed.
One may further note that $w_{\tau\tau}$ is of the same order of magnitude as $w$.\footnote{Recall that we are using the time scale of the excitation, $\tau=\Omega t$.}
Using the values for $\hat w(s)/L$ and $\hat w_\xi(s)/L$ from \tref{signaturemoveparams}, one obtains an estimate for the order of magnitude of $p_{\tau\tau}^{\mrm{{slope}}}$:
\begin{itemize}
\item[Case 1:] $6\cdot 10^{-6}$,
\item[Case 2:] $1\cdot 10^{-4}$, and
\item[Case 3:] $3\cdot 10^{-4}$.
\end{itemize}
%
To estimate the order of magnitude of the rocking-induced acceleration $p_{\tau\tau}^{\mrm{{rock}}}$ in \eref{sloperot}, the approximation $p\approx s$ is also used in the argument of $w$ and its derivatives (it was checked that this does not change the estimated order of magnitude).
This yields the estimation:
\begin{itemize}
\item[Case 1:] $2\cdot 10^{-4}$,
\item[Case 2:] $1\cdot 10^{-3}$, and
\item[Case 3:] $2\cdot 10^{-3}$.
\end{itemize}
Apparently, the individual contributions to the rocking-induced acceleration are two (Case 1) or one order of magnitude larger (Case 2 and 3) than the slope-induced acceleration.
As mentioned before, however, the individual contributions differ by their sign, so that the net contribution is much smaller.
The net contribution is generally non-zero because of the asymmetry.
More specifically, there is a left-right asymmetry ($w$ and its derivatives are actually to be evaluated at $s^{(\mrm{l})}\neq s^{(\mrm{r})}$), and a top-down asymmetry ($\ell$ and $|\sigma|$ differ between upper and lower contacts).
In fact, it will turn out that both asymmetries are needed for a non-zero net contribution.
Further, it is important to note that the rotational accelerations are active only when one contact is closed.
This is in contrast to the slope-induced acceleration, which is also active when two contacts are closed.
\\
To determine the net contribution of slope- and rocking-induced acceleration, the increment of the horizontal slider position per excitation cycle, $\Delta s$, is estimated.
This permits also the comparison with the pitching locomotion.
Formally, the slip per cycle is obtained by integration
\ea{
\Delta s^{\mrm{slip}} &=& \int\limits_{0}^{2\pi} ~\left( p_{\tau}(0) + \int\limits_{0}^{\tau} p_{\tau\tau}(\tau^*)\dd\tau^*\right)~\dd\tau\fk \label{eq:dsslip} \\
&=& \Delta s^{\mrm{slope}} + \Delta s^{\mrm{rock}} \nonumber
}
As the rocking-induced acceleration has larger magnitude and the sign oscillates (from contact to contact), there will be phases of zero velocity, $p_\tau=0$, in every cycle.
It is thus assumed that the formally introduced initial velocity term, $p_{\tau}(0)$, has negligible effect on $\Delta s^{\mrm{slip}}$.
\\
For simplicity, it is assumed that the slope-induced acceleration is active during the whole period.
Further, the approximation $p\approx s$ is deemed appropriate in period-average for the slope-induced acceleration; \ie, $w^{\mrm P}\approx w(s)$ and the analog for the derivatives.
Consequently, one can use $w(s)=\hat w(s)\cos\tau$, $w_{\tau\tau}(s) = -\hat w(s)\cos\tau$, and $w_{\xi}=\hat w_\xi(s)\cos\tau$.
This way, one obtains the approximation:
\ea{
\Delta s^{\mrm{slope}} &=& \frac{\hat w(s)}{L} \frac{\hat w_\xi}{L} \int\limits_{0}^{2\pi} \int\limits_{0}^{\tau} \cos^2\tau^* \dd\tau\fk \nonumber \\
&=& \pi^2 \frac{\hat w(s)}{L} \frac{\hat w_\xi}{L}\fp \label{eq:dsslope}
}
For the different cases, this term evaluates to:
\begin{itemize}
\item[Case 1:] $6\cdot 10^{-5}$,
\item[Case 2:] $1.2\cdot 10^{-3}$, and
\item[Case 3:] $2.6\cdot 10^{-3}$.
\end{itemize}
In Case 1, the term is of similar order of magnitude as the pitching-induced locomotion per cycle ($|\Delta s^{\mrm{pitch}}| = 7.5\cdot 10^{-5}$, \cf~\eref{DeltaXQ}).
Recall, however, that free sliding was assumed; it is still to be analyzed if the horizontal accelerations are sufficiently large to overcome dry friction.
\\
Next, the rocking-induced slip, $\Delta s^{\mrm{rock}}$, is estimated.
To integrate $p_{\tau\tau}^{\mrm{rock}}$ in \eref{dsslip}, a certain contact sequence must be postulated; \ie, when does which contact close and when does it open.
As mentioned before, the lower contacts tend to be closed during the half period with $w>0$, and the upper contacts tend to be closed during the half period with $w<0$.
Also recall that the rocking-induced acceleration is only active when a single contact is closed.
An upper estimate of $\Delta s^{\mrm{rock}}$ is obtained by assuming that each of the four contacts is closed exactly for a quarter period.
It can be shown that it is not relevant if first the left contact is closed for a quarter period, and then the right, or the other way around.
With this, one obtains:
\ea{
\Delta s^{\mrm{rock}} &=& \sum\limits_{i=1}^4
-\zeta_{z,i}\frac{\hat w(s+\zeta_{x,i}\frac{B}{2L})}{L} \frac{m\ell_i^2}{J^{(C)}+m\ell_i^2}\frac{\sin\left(2\sigma_i\right)}{2}\fk \nonumber\\
&=& -\left( \frac{\hat w(s+\frac{B}{2L})}{L} - \frac{\hat w(s-\frac{B}{2L})}{L} \right)\left(\Lambda^{\mrm{o}}-\Lambda^{\mrm u}\right)\fk \label{eq:dsrot} \\
\Lambda^{\mrm{o}} &=& \frac{m\ell_1^2}{J^{(C)}+m\ell_1^2} \frac{\sin\left(2|\sigma_1|\right)}{2}\fk \label{eq:LambdaO}\\
\Lambda^{\mrm{u}} &=& \frac{m\ell_2^2}{J^{(C)}+m\ell_2^2} \frac{\sin\left(2|\sigma_2|\right)}{2}\fp \label{eq:LambdaU}
}
Note that $\ell_1=\ell_4$, $\ell_2=\ell_3$ and $|\sigma_1|=|\sigma_4|$, $|\sigma_2|=|\sigma_3|$ (\tref{locoparams}).
For the different cases, $\Delta s^{\mrm{rock}}$ evaluates to:
\begin{itemize}
\item[Case 1:] $-2.4\cdot 10^{-5}$,
\item[Case 2:] $-1.1\cdot 10^{-4}$, and
\item[Case 3:] $-1.4\cdot 10^{-4}$.
\end{itemize}
In Case 1, $|\Delta s^{\mrm{rock}}|$ is slightly smaller than $|\Delta s^{\mrm{slope}}|$; those two terms partially cancel.
In Case 2 and Case 3, $|\Delta s^{\mrm{slope}}|$ is an order of magnitude larger than $|\Delta s^{\mrm{rock}}|$.
It should be added that the actual rocking-induced slip is even smaller because \eref{dsrot} ignores phases with two closed contacts which inhibit rocking-induced acceleration.

\subsubsection{Sticking vs. sliding}
The question still needs to be addressed \MOD{under what conditions} the horizontal accelerations are sufficient to overcome static dry friction.
To this end, the ratio between tangential (horizontal) and normal (vertical) acceleration is determined, and this ratio is compared to the identified friction coefficient of $0.2$ (\tref{params}).
The vertical acceleration is dominated by $w_{\tau\tau}$ in the largest part of the high-amplitude branch.
In this case, the analysis is simple:
The approximation of the horizontal acceleration is given in \eref{sloperot}, and one can see that both terms, $p_{\tau\tau}^{\mrm{rock}}$ and $p_{\tau\tau}^{\mrm{slope}}$ are proportional to $w_{\tau\tau}$.
For $p_{\tau\tau}^{\mrm{slope}}$, the coefficient is the beam's slope, and since $|w_\xi^{\mrm P}/L|<0.035\ll 0.2$ (\cf~\tref{signaturemoveparams}), slope-induced acceleration \emph{alone} cannot be expected to overcome static dry friction \MOD{ with the friction coefficient of $0.2$ here.
In fact, an extremely low effective friction coefficient would be needed for this, which was achieved by roller bearings in \cite{Miranda.1998}.
}
On the other hand, $p_{\tau\tau}^{\mrm{rock}}$ is also proportional to $w_{\tau\tau}/L$, where the proportionality coefficient is $\Lambda^{\mrm{o}}$ or $\Lambda^{\mrm{u}}$.
Interestingly, $\Lambda^{\mrm{o}}$ and $\Lambda^{\mrm{u}}$ depend only on geometric and inertia properties, but not on the vibration level.
The values are $\Lambda^{\mrm{o}}=0.20$ and $\Lambda^{\mrm{u}}=0.17$, which is very close to the identified friction coefficient of $0.2$.
On the high-amplitude branch, it is expected that the rocking-induced acceleration can induce some sliding friction.
When sliding towards the beam's center occurs, the slope-induced horizontal acceleration speeds up the slider, and it slows the slider down in phases where rocking-induced sliding occurs away from the beam's center.
In this sense, the \emph{rocking-induced acceleration is an enabler for the slope-induced acceleration} (which by itself would not overcome static friction).
Analyzing the parameters $\Lambda^{\mrm{o}}$ and $\Lambda^{\mrm{u}}$ further, one finds an optimal pendulum angle of $|\sigma|=\pi/4$ for which $\sin(2|\sigma|)/2$ assumes its maximum of $0.5$.
Comparing with \tref{locoparams}, one finds that the design, which did not rely on the theory developed in the present work, was close to optimal in this respect.
From this theory, it is clear that introducing a vertical distance $d\neq 0$ between contact center and center of mass facilitates the locomotion:
For $d=0$, $\ell\approx B/2$, $|\sigma|\approx \pi/2$, so that $\sin(2|\sigma|)/2\approx 0$ and $\Lambda^{\mrm{o}} \approx 0 \approx \Lambda^{\mrm{u}}$.
\\
In some parts of the low-amplitude branch, specifically at relatively small vibration levels, besides the elastic acceleration $w_{\tau\tau}$, the base acceleration $w_{\tau\tau}^{(0)}$ and the gravity acceleration are also relevant.
It is believed that this mitigates sliding to some extent (sticking at upper contacts; lower effective acceleration because $w$ and $w^{(0)}$ are in anti-phase).
Thus, it is expected that the slope- and the rocking-induced slip are considerably smaller than their estimates given in \erefs{dsslip}-\erefo{dsrot} on the low-amplitude branch.
As a consequence, the pitching cycle dominates the locomotion, leading to slider transport away from the beam's center.

\section{Locomotion: Validation of theory and further observations\label{sec:locovalidation}}
In this section, the simplified theory on the locomotion is compared against numerical simulation results obtained from the detailed model (experimentally validated in \cite{Muller.2023}).
Recall that the numerical simulations are carried out using the time integration procedure described in \cite{Muller.2023}, where unilateral interactions in the contact normal direction and Coulomb dry friction in the tangential direction are modeled at the four possible contact points (\fref{SAS}b).
\MOD{The simulation code is available online via \url{https://github.com/maltekrack/simSAS}.}
The results are grouped by the different phases / representative cases.
The Pseudo-Constrained Slider (PCS) model is used in the simulation, which is described in \ssref{sSIMvalidation}.
Compared to the Free Slider (FS) model, this facilitates obtaining statistically converged metrics of the chaotic behavior for given slider locations.
The consistency of PCS and FS model was already shown with regard to the \sSIM in \cite{Muller.2020,Muller.2023}.
In \fref{locomotion_temp_separable}b the consistency with regard to the locomotion speed is demonstrated.
It is interesting to note that the results were inconsistent for the previously proposed constrained slider model \cite{Krack.2017b}.
The vertical guide used in that case was found to be too intrusive, which is plausible in view of the forms of locomotion proposed in \sref{locotheory}.
This was the main motivation for the development of the PCS model.
\\
It is useful to recall the most important simplifications made in \srefs{sSIM}-\srefo{locotheory}:
\begin{itemize}
    \item The beam vibrates harmonically with constant amplitude in a single mode shape. That mode shape was obtained for idealized boundary conditions and without the effect of the slider.
    \item Simple contact sequences were \MOD{assumed, neglecting free-flight and sliding or sticking phases.}
\end{itemize}

\subsection{Phase 1: Locomotion away from the beam's center on the low-amplitude branch}
For Case 1 (representative point in Phase 1), results obtained using the PCS model are shown in \fref{overview_s027_down}.
\MOD{
The beam's response shows a strong contribution in the fundamental harmonic of the excitation.
}
The elastic displacement is in good approximation in anti-phase with the base displacement.
This is in agreement with the theory postulated on the \sSIM in \sref{sSIM}.
Further, Case 1 shows the lowest level of beam vibration among the three cases, and the vibration is strongly modulated, as explained by the analogy of the self-adaptive system to a host structure with an impact absorber.
The slope angle $\alpha = -\operatorname{atan}(w_\xi(s)/L) \approx -w_\xi(s)/L$ is in the range of $\hat{\beta}_\mrm{rel}$, as predicted for Case 1 (\tref{signaturemoveparams}).
\MOD{
The frequency spectrum of the beam's response is shown in \fref{FFT}.
Besides the resonant harmonic, frequency $\Omega$, a high peak occurs at a slightly lower frequency.
This corresponds to the system's effective natural frequency $\tilde\omega$.
The relative frequency difference is $(\Omega-\tilde\omega)/\Omega\approx0.1$.
This explains why the modulation is approximately periodic with ca.~10 excitation periods.
It should be remarked that this modulation frequency, predicted by the PCS model, is in good agreement with the simulation results of the FS model, and the experimental results \cite{Muller.2023}.
}
\onelargefig{.9}
{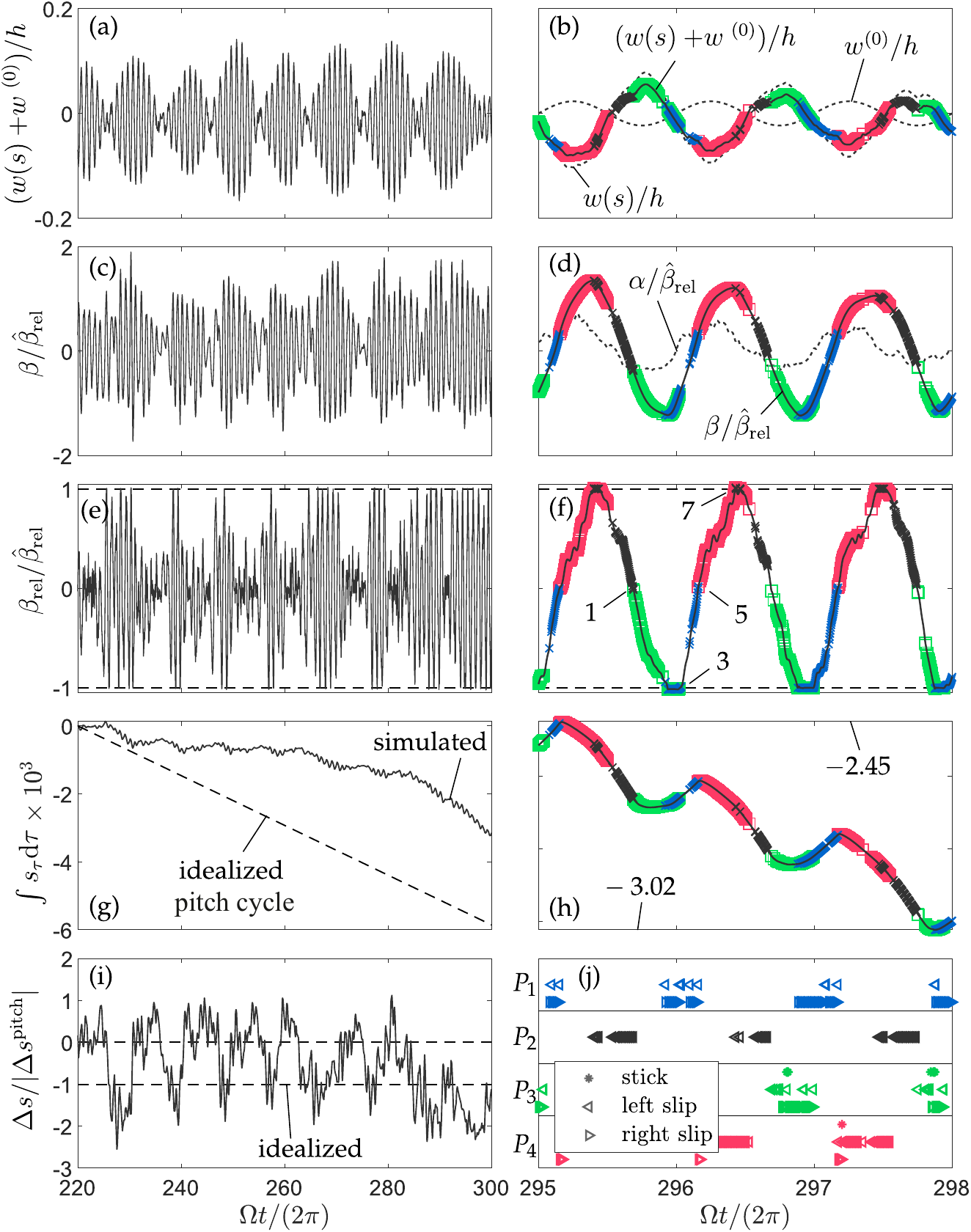}
{Representative situation (\emph{Case 1}) showing dynamics with slider transport away from the beam's center; low-amplitude branch at $s=0.27$: (a) long term behavior of the beam's total displacement, (c) total slider rotation, (e) relative slider rotation, (g) horizontal slider position, (b, d, f, h) zoom into the graph on the left, markers according to \fref{allocation_cts_pts} indicate closed contacts, (i) horizontal slider transport per cycle, (j) contact states.}
{overview_s027_down}
\onefig{1}{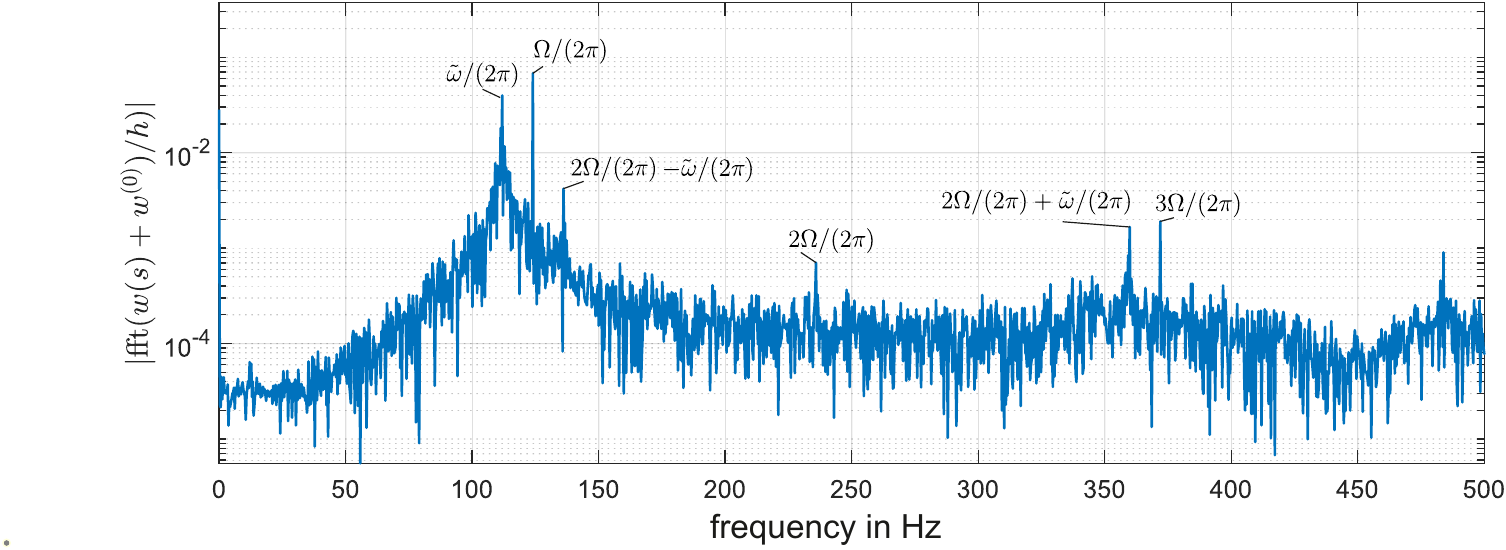}{Frequency spectrum of the long-term behavior of the beam's total displacement (\emph{Case 1}), corresponding to \fref{overview_s027_down}a.}{FFT}
\\
The pitching cycle can be clearly identified:
The pitch limit is reached frequently; the contact sequence and the relative rotation show a good correlation with the sketch of the idealized cycle in \fref{visualization_idealized}.
Note that the closed contacts are indicated by markers following the labeling defined in \fref{allocation_cts_pts}.
The average horizontal slider velocity is about factor two slower than in the idealized case.
This is attributed to the modulation of the beam's vibration amplitude.
The pitching cycle occurs only for a few cycles near the maximum of the amplitude envelope.
This slows down the average locomotion.
In \fref{overview_s027_down}i, the actual locomotion $\Delta s$ per cycle is shown, normalized by the idealized value from the pitch cycle, $\Delta s^{\mrm{pitch}}$ ($\Delta s/|\Delta s^\mrm{pitch}|=-1$ means perfect match; the negative sign comes from the fact that for $s<0.5$, the pitching cycle transports to the left/away from the beam's center).
Apparently, the pitching locomotion is actually faster in some time spans than the idealized one.
This is because the contacts are actually sliding, in contrast to what was assumed in \ssref{pitchingCycle}.
In fact, sticking is barely observed (\fref{overview_s027_down}j).
When the contacts $\Ptwo$ or $\Pfour$ are closed, rocking-induced sliding accelerates the locomotion.
In contrast, when the contacts $\Pone$ or $\Pthree$ are closed, the rocking-induced acceleration slows down the locomotion, leading to an approximately constant position of the contact center.
The sliding directions are largely consistent with those predicted in \ssref{rotindsl}.
Recall that rocking-induced sliding has a net component away from the beam's center (for an in-average symmetric contact sequence).
The resulting slider position is the superposition of a monotonic linear function (pitch-induced locomotion occurring at approximately constant speed) and an oscillation (rocking-induced sliding in alternating directions).
In the time spans with low beam amplitude, the contact pattern is irregular.
In the absence of the pitching cycle here, some rotation- and slope-induced sliding occurs, with a small net component (small because the components partly cancel each other, and because of the lower vibration level), as predicted, pointing towards the beam's center.
In \cite{Shin.2020}, interestingly, it was conjectured that the pitch limit is never reached in Phase 1 where the slider moves away from the beam's center.
The present results falsify this conjecture.
\\
As announced in \ssref{pitchingCycle}, the pitching cycle is now interpreted from a kinetic point of view.
This is important because, in principle, the exact mirror of the process sketched in \fref{visualization_idealized} is also kinematically possible and would lead to locomotion in the opposite direction (towards the beam's center).
The key question is: What causes the relative rotation, and what phase between this relative rotation and the beam displacement is to be expected?
To answer this, suppose initially no relative rotation between beam and slider.
Further, suppose that the upper contacts are closed initially and the beam displacement and velocity are $w=0$ and $w_{\tau}>0$, respectively. 
As explained before, the transverse acceleration $w_{\tau\tau}$ is also zero at this point and becomes negative.
Consequently, the upper contacts open.
In this free-flight phase, the slider moves with approximately constant velocity in vertical direction.
As the beam decelerates, the lower contacts close at some point.
Due to the clearance, this does not happen instantaneously.
In the meantime, the beam has rotated (counter-clockwise in the left half).
Consequently, the lower left contact ($\Ptwo$) closes first (state 8 of pitching cycle in \fref{visualization_idealized}).
Then, the slider rotates until the lower right contact ($\Pthree$) closes (state 9=state 1).
Due to its rotational inertia, the slider rotates further, so that $\Ptwo$ opens (state 2).
If the rotational velocity is sufficient, the rotation continues until the upper left contact ($\Pone$) closes (state 3).
In the meantime, the beam passes through $w=0$ with $w_\tau<0$, and $w_{\tau\tau}$ becoming positive so that the lower contact ($\Pthree$ here) opens (state 4).
Then the process continues with its mirror (rotation around $\Pone$ until $\Pfour$ closes (state 5); rotation around $\Pfour$ (state 7) until $\Ptwo$ closes (state 8)).
With this, the pitch limit is expected to be reached near $w=0$, while the situation of zero relative rotation (where either the two upper or the two lower contacts are closed) occurs near the reversal points, $w=\pm \hat w$, which implies a phase shift of approximately $\pi/2$ between beam displacement and relative rotation (beam-slider).
This is confirmed by the results shown in \fref{overview_s027_down}b and f.

\subsection{Phase 2: Locomotion towards the beam's center on the high-amplitude branch}
The results for Phase/Case 2 are shown in \fref{overview_s027_up}.
Here, the beam's response is of much higher level, and strongly dominated by the resonant harmonic.
The resulting vibration is almost periodic.
The base displacement is in phase with the elastic one, and the elastic displacement clearly dominates.
Again, this is in agreement with the theory postulated on the \sSIM in \sref{sSIM}.
\onelargefig{1}
{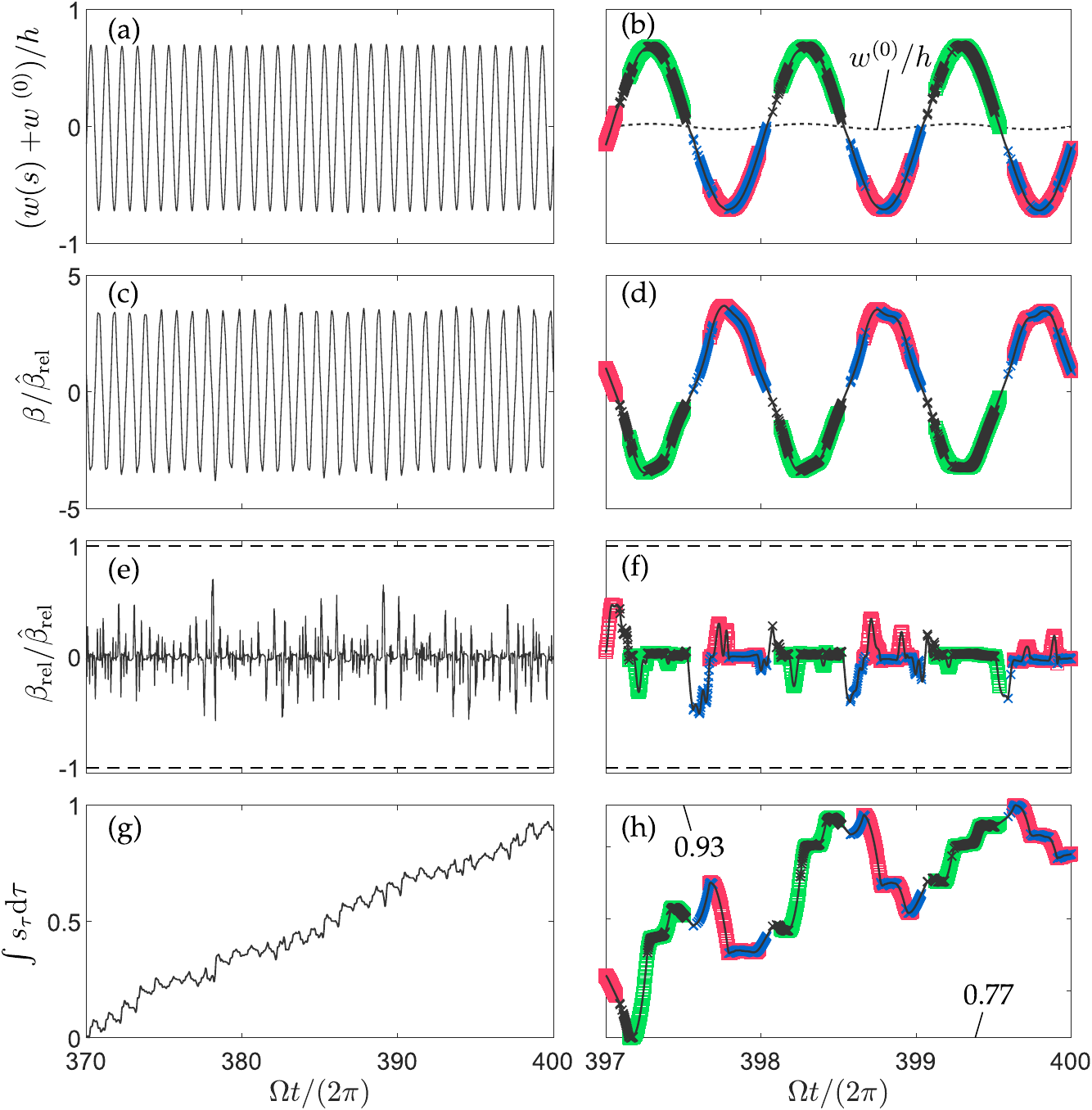}
{Representative situation (\emph{Case 2}) showing dynamics with slider transport towards the beam's center; high-amplitude branch at $s=0.27$: (a) long term behavior of the beam's total displacement, (c) total slider rotation, (e) relative slider rotation, (g) horizontal slider position, (b, d, f, h) zoom into the graph on the left, markers according to \fref{allocation_cts_pts} indicate closed contacts.} 
{overview_s027_up}
\\
In contrast to Case 1, the pitch limit is never reached.
This is attributed to the fact that the clearance is small compared to the vibration level.
This means there is much less delay between the opening of upper/lower contacts and the closure of the respective other.
In this relatively short time span, the beam has less time to rotate significantly.
Apparently, other sources of asymmetry that could give rise to some intermediate rotations, are decelerated by the sliding-pendulum effect.
Thus, there is less relative rotation between slider and beam, so that the pitch limit is not reached.
Instead, the slider follows closely the displacement and rotation of the beam.
Most of the time, both contacts are closed on either side (\emph{plunging motion}).
Some free flight occurs near $w\approx 0$, and some time spans with only one closed contact are present.
\\
Interestingly, there is considerable left-right asymmetry in the contact sequence.
This is attributed to the finite width of the slider and the fact that the beam has higher vibration amplitude closer to the center.
As indicated in \tref{signaturemoveparams}, the right contacts see a more than $40~\%$ larger displacement amplitude than the left contacts.
For simple kinematic reasons, this leads to longer contact times on the right; also, due to the higher transverse acceleration, the right contacts are exposed to higher normal contact forces.
The left contacts mainly follow, and close because of the sliding-pendulum effect.
In fact, the right contacts only close and open once per half cycle, while the left have intermediate liftoff-collision events.
\\
Assuming sliding without friction, the slope-induced slip per cycle was estimated as $\Delta s^{\mrm{slope}} \approx 1.2\cdot 10^{-3}$ (\ssref{rotindsl}).
Apparently, the actual $\Delta s$ is by a factor of about 30 smaller (in average).
Hence, the slope-induced locomotion cannot be active all the time, in contrast to what was assumed for \eref{dsslip}.
Apparently, most horizontal transport occurs when only the lower right ($\Pthree$) or the upper right ($\Pfour$) contact is closed.
Here, sliding is induced by rocking.
This is also supported by the sliding directions shown in \fref{detials_transport_to_center}c, which are largely consistent with those predicted in \ssref{rotindsl}.
According to the theory, the contributions of the different contacts to the rocking-induced slip produce a net component that is at least one order of magnitude smaller than the slope-induced slip.
As explained earlier, the high rocking-induced accelerations cause the contacts to slide and hence enable slope-induced accelerations to take effect.
In time spans with two contacts closed on one side, the horizontal velocity is very small.
In that situation, typically one point sticks, or sliding occurs in opposite directions.
The sliding in opposite directions seems surprising on first sight.
It is an inevitable consequence of the fact that the beam stretches under the rigid slider.
\onefig{.49}
{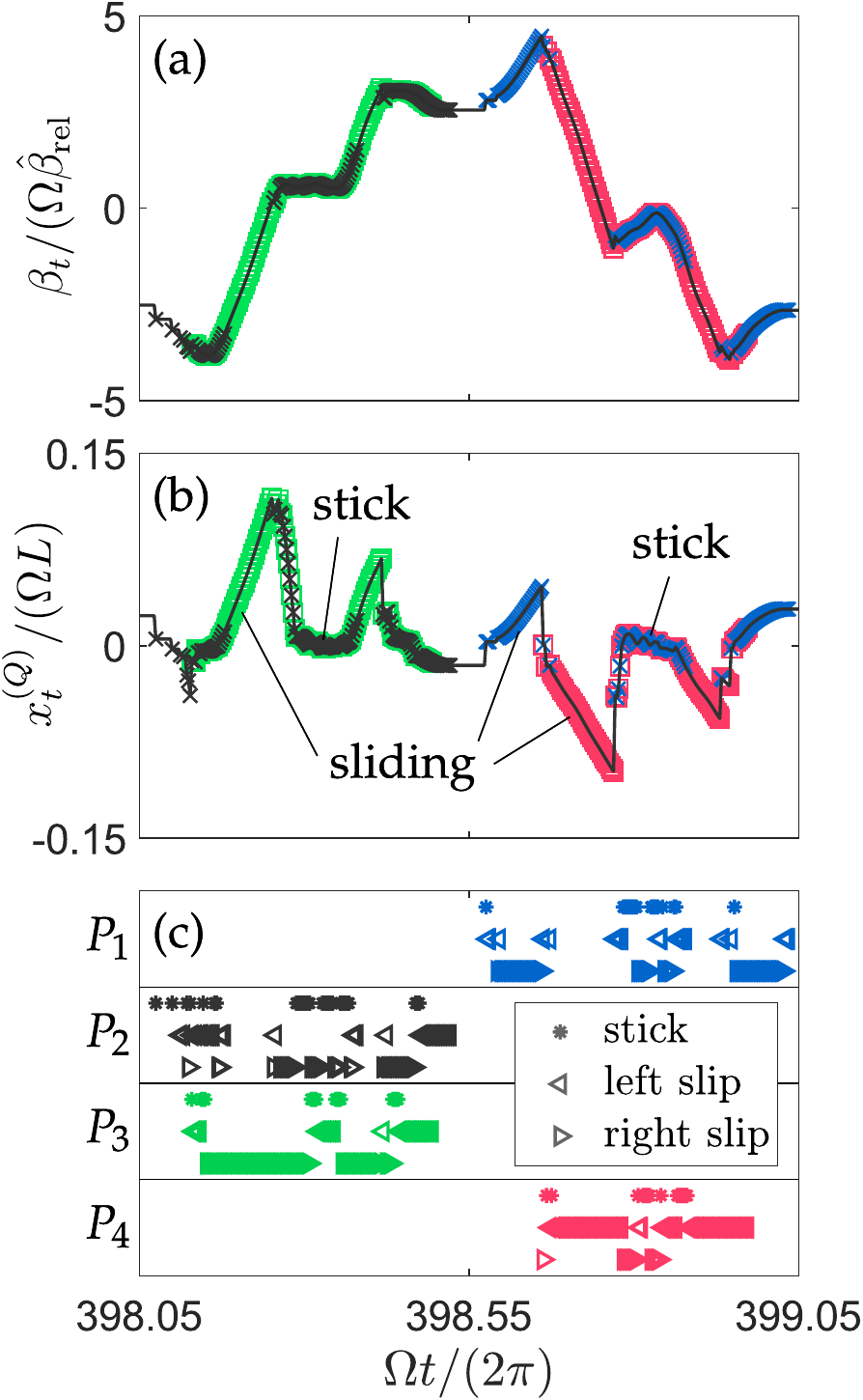}
{Representative period of Phase 2: (a) slider's total rotational velocity, (b) horizontal slider velocity, (c) contact states.}
{detials_transport_to_center}
%

\subsection{Phase 3: Zero effective locomotion on the high-amplitude branch}
The results for Phase/Case 3 are shown in \fref{overview_s0328_stst}.
As in Case 2, the beam displacement is almost periodic and strongly dominated by the resonant harmonic of the vibration.
The contact sequence is also in good approximation periodic.
Interestingly, the pitch limit is reached again, twice per excitation period.
It should be emphasized that the beam's displacement amplitude is about $80~\%$ larger than in Case 2.
The left-right asymmetry (deflection shape) and the rotational inertia inevitably induce some relative rotations.
In combination with the higher energy in the system, the pitch limit is reached.
%
\onelargefig{1}
{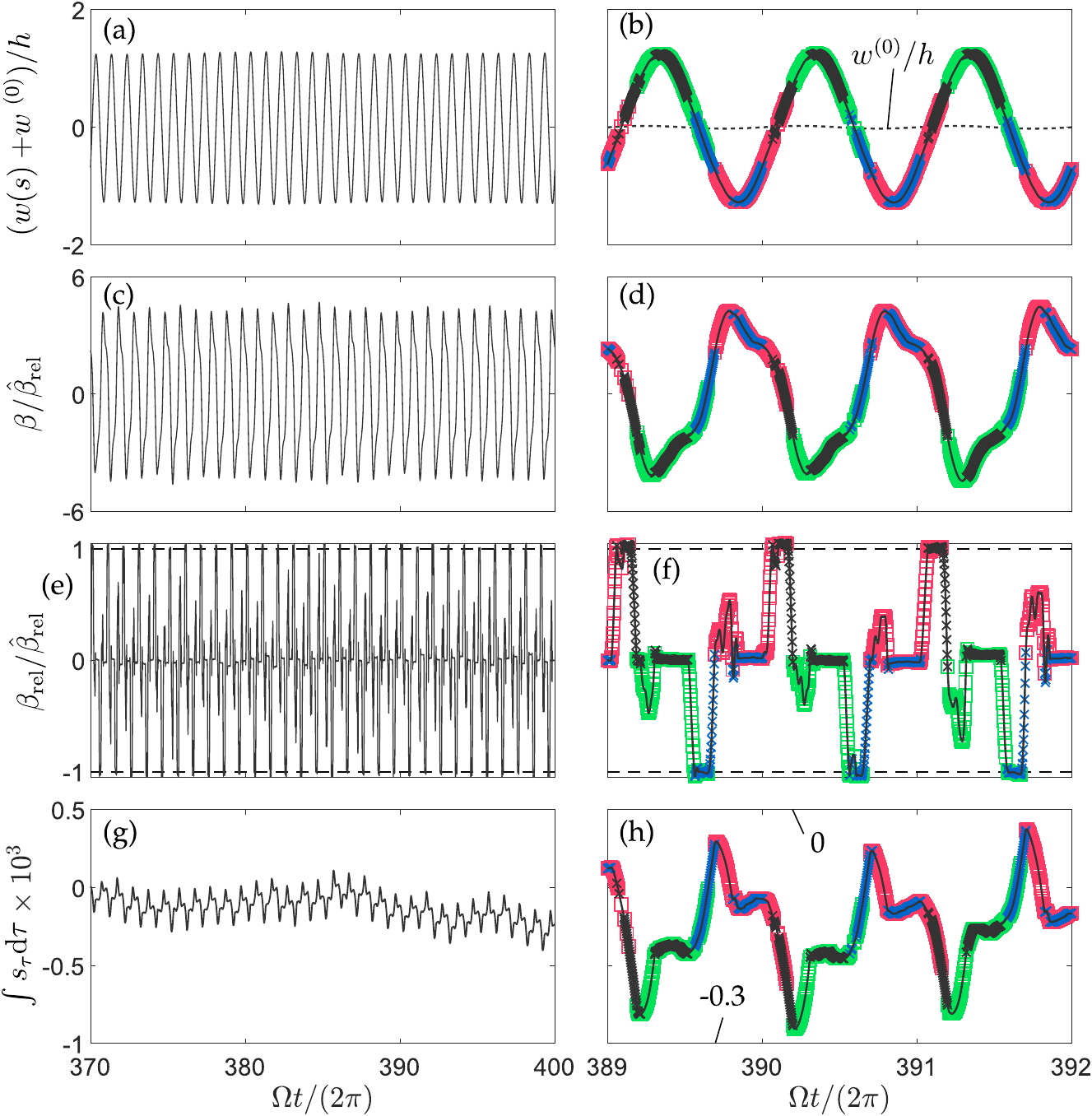}
{Representative situation (\emph{Case 3}) showing dynamics with negligible slider transport; high-amplitude branch at $s=0.328$: (a) long term behavior of the beam's total displacement, (c) total slider rotation, (e) relative slider rotation, (g) horizontal slider position, (b, d, f, h) zoom into the graph on the left, markers according to \fref{allocation_cts_pts} indicate closed contacts.}
{overview_s0328_stst}
\\
The slider dynamics depicted in \fref{overview_s0328_stst} has time spans that resemble the pitching cycle, and time spans that resemble the plunging motion in Case 2.
In Case 2, free flight occurs at the transitions between the time spans with only upper/lower contacts closed.
In Case 3, instead of free flight, the pitching sequence (state 1 and state 5 of the idealized cycle in \fref{visualization_idealized}) takes place.
Apparently, both pitching-, as well as rotation- and slope-induced locomotion occur.
As one moves towards the beam's center along the high-amplitude branch, the system energy and, specifically, the slider's rotational energy increases.
Consequently, the pitching-induced locomotion becomes more important there.
At some point, pitching-induced and rocking-induced locomotion cancel each other (\fref{locomotion_temp_separable}b).
This is a state of dynamic equilibrium.
In previous works \cite{Miller.2013,Shin.2020}, it was conjectured that the slider stops due to dry friction.
The present results falsify that conjecture.
\onefig{1}
{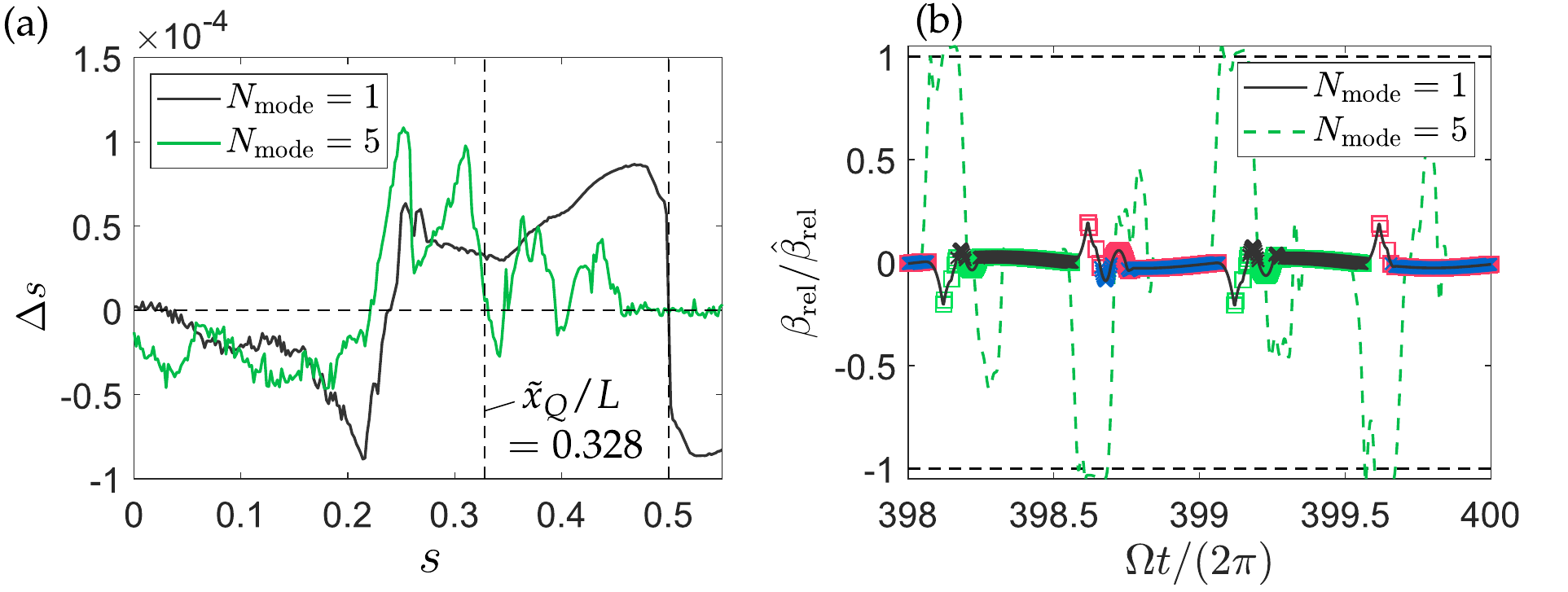}
{Comparison of the locomotion on the high-amplitude branch for modal truncation order $\nmod=1$ and $\nmod=5$: (a) horizontal slider transport per period, (b) slider's rotation relative to the beam at $s=0.328$.}
{transport_HAB_N1_N5}
\\
In \fref{transport_HAB_N1_N5}, simulation results are compared for two different models, one where the beam is truncated to a single mode, and one where the beam is truncated to five modes (which was found sufficient to capture the locomotion accurately).
The results show qualitative differences:
For $\nmod=1$, the pitching limit is never reached on the high-amplitude branch, so that the slider would never stop before reaching the beam's center (in contrast to what was observed in the experiment).
Apparently, the local flexibility of the contact region, and to some extent also the higher-frequency dynamics, are crucial for properly describing the locomotion here.
In \frefsf{transport_HAB_N1_N5}{b}, the relative slider rotation is shown for $s=0.328$, where the slider is in dynamic equilibrium according to the PCS/FS model (Case 3).
For $\nmod=1$, a relatively regular plunging motion is observed where both contacts on either side are closed most of the time.
The relative rotation is close to zero most of the time, except for small kinks at the beginning of the plunging phases.
The second opening of the left contact that initiates the relative rotation to the pitch limit is never observed.
It should be emphasized here that the local elasticity is crucial for initiating the pitching locomotion in Phase 3.
On the low-amplitude branch (Phase 1), in contrast, it was found that good qualitative and quantitative agreement can be achieved with $\nmod=1$.
%

\section{Conclusions} \label{sec:conc}
The qualitative behavior of the self-adaptive system can be largely described using a single-mode model of the beam, and assuming \MOD{(mono-frequency)} harmonic vibrations.
One exception is the stopping of the slider on the high-amplitude branch, for which the local flexibility of the beam must be properly described. 
The strong amplitude modulation at low vibration levels, as opposed to the almost periodic behavior at high vibration levels, is not surprising when noticing the similarity to a host structure with impact absorber, for which these regimes are well-known.
On the low-amplitude branch, the locomotion away from the beam's center is driven by the described pitching cycle.
The required relative rotation with the appropriate phase to the beam's vibration is induced by the left-right asymmetry (left and right contacts see a different vibration amplitude due to the beam's modal deflection shape in connection with the finite slider width) and the slider's rotary inertia.
On the high-amplitude branch, the rocking-induced accelerations (sliding-pendulum effect), which occur when one contact is closed, facilitate sliding, and thus enable the slope-induced accelerations to take effect.
The slope-induced accelerations alone would not be able to overcome static friction.
To reach sufficient rocking-induced acceleration, it is crucial to have a certain vertical distance between the slider's center of mass and the contact center.
Also, the clearance is crucial: Without it, the pitching cycle and the rocking-induced sliding are impossible (both of which require relative rotation between slider and beam).
The slider stops not because of dry friction, but because there is dynamic equilibrium between the two described forms of locomotion.
\\
In the future, it would be interesting to experimentally validate the theoretical findings of the present work, \eg using high-speed camera footage of the locomotion (the slider side would have to be partially open or transparent).
Also, we are convinced that the knowledge gained in the present work on the locomotion and on the super-slow system dynamics is crucial for making the self-adaption quicker and more robust to excitation and initial conditions.


\end{document}